\documentclass{article}

\usepackage{PRIMEarxiv}

\usepackage[utf8]{inputenc} 
\usepackage[T1]{fontenc}    
\usepackage{hyperref}       
\usepackage{url}            
\usepackage{booktabs}       
\usepackage{amsfonts}       
\usepackage{nicefrac}       
\usepackage{microtype}      
\usepackage{lipsum}
\usepackage{fancyhdr}       
\usepackage{graphicx}       
\usepackage[linewidth=0.5pt]{mdframed}
\usepackage{algorithm} 
\usepackage{algorithmic}

\usepackage{graphicx}
\usepackage{caption}

\graphicspath{{media/}}     

\usepackage{amsmath,amsfonts,bm}

\def\eqref#1{equation~\ref{#1}}

\def\1{\bm{1}}

\DeclareMathAlphabet{\mathsfit}{\encodingdefault}{\sfdefault}{m}{sl}
\SetMathAlphabet{\mathsfit}{bold}{\encodingdefault}{\sfdefault}{bx}{n}

\DeclareCaptionLabelFormat{nocaption}{} 
\title{Personalized Federated Learning on Data with Dynamic Heterogeneity under Limited Storage}

\author{
  Sixing Tan\thanks{First author} \\
  Faculty of Computing \\
  Harbin Institute of Technology \\
  \texttt{hit\_tsx@163.com}\\
   \And
  Xianmin Liu\thanks{Corresponding author}\\
  Faculty of Computing \\
  Harbin Institute of Technology \\
  \texttt{liuxianmin@hit.edu.cn}\\
}

\begin{document}
\maketitle

\begin{abstract}
  Recently, a large number of data sources opened up by informatization intensify the data heterogeneity, the faster speed of data generation and the gradual implementation of data regulations limit the storage time of data. In personalized Federated Learning (pFL), clients train customized models to meet their personal objectives. However, due to the time-varying local data heterogeneity and the inaccessibility of previous data, existing pFL methods not only fail to solve the catastrophic forgetting of local models, but also difficult to estimate the degree of collaboration between clients. To address this issue, our core idea is a low consumption and high-quality generative replay architecture. Specifically, we decouple the generator by category to reduce the generation error of each category while mitigating catastrophic forgetting, use local model to improving the quality of generated data and reducing the update frequency of generator, and propose a local data reconstruction scheme to reduce data generation while adjusting the proportion of data categories. Based on above, we propose our pFL framework, pFedGRP, to achieve personalized aggregation and local knowledge transfer. Comprehensive experiments on five datasets with multiple settings show the superiority of pFedGRP over eight baseline methods.
\end{abstract}

\section{Introduction}
\begin{figure}[ht]
  \vskip 0.2in
  \begin{center}
  \label{Figure 1}
  \centerline{\includegraphics[width=0.6\columnwidth]{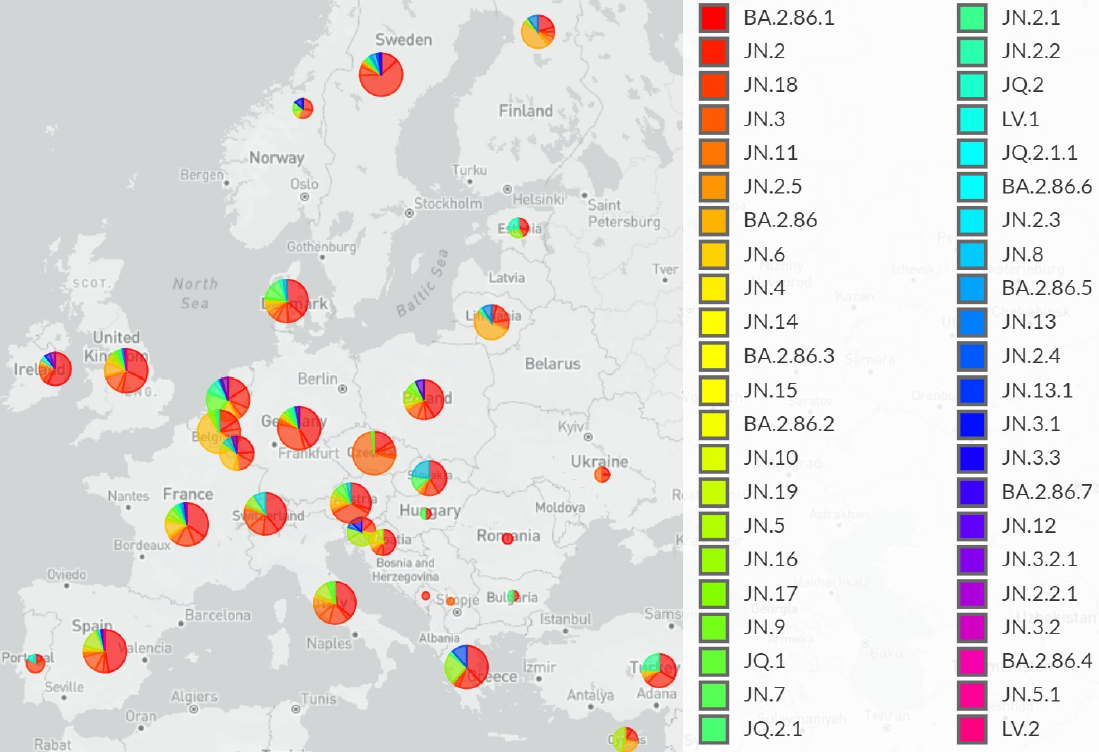}}
  \caption{\text{\small\sl Figure 1.} The proportion of different types of the COVID\-19 virus in various regions of Europe in January 2025. The data is sourced from \url{https://gisaid.org/hcov19-variants/}.}
  \end{center}
  \vskip -0.2in
\end{figure}

Federated Learning (FL) \cite{01} is an emerging distributed machine learning framework with privacy protection. In the forbidden of transmitting local dataset, clients collaborate to train a shared global model by transmitting the updates of the local models.
However, in practice, data heterogeneity within and between clients varies over time \cite{02}, and the accessible data on the client side is often limited by relevant data regulations and policies \cite{N03} \cite{04}.
For example, health institutions in different regions can use FL to conduct research on COVID-19 \cite{05} together, but the high mutation speed of the virus can lead to differences in the distribution and trends of medical data across institutions (see Fig \hyperref[Figure 1]{1}), and the data protection regulations \cite{N03} limit the storage time for original data.
We denote the FL situation above as “Data with Dynamic Heterogeneity under Limited Storage”. In this situation, the global model is often difficult to meet the utility of each client \cite{06} \cite{07}, FL should customize personalized global model for clients to adapt to their local data, this type of FL is denoted as personalized Federated Learning (pFL).

The fundamental challenge in pFL lies in estimating the data heterogeneity between clients to tradeoff the individual utilities and collaborative benefits. Specifically, the similar gradients can improve the generalization of models \cite{08}, and the similarity of the local updates is inversely proportional to the data heterogeneity between clients \cite{10}, meaning that collaboration will bring less benefits to clients under higher data heterogeneity.
To estimate data heterogeneity between clients, existing pFL works, such as Ditto \cite{11}, FedRep \cite{12}, KT-pFL \cite{12}, propose different methods from multiple perspectives including estimating model distance, partial aggregation and knowledge transfer.
However, existing pFL works typically estimate the data heterogeneity through the information of local models, making it difficult to focus on the performance of the model on the inaccessible previous data, known as catastrophic forgetting \cite{14} \cite{15}. Thereby, the personalized global model obtained by the client may not necessarily meet its requirements \cite{16}.
Moreover, in reality, clients may meet the data that other clients have already encountered, but under higher data heterogeneity, the personalized global model contains less global information, thereby reducing the generalization of the model on those data \cite{17}.

Inspired by Continuous Learning (CL) through generated replay \cite{18} \cite{19}, we attempt to combine pFL with generated replay to achieve personalized aggregation, alleviating catastrophic forgetting and improving model generalization.
Although there are already many Federated Continuous Learning (FCL) works such as FedCIL \cite{20}, CFeD \cite{21}, TARGET \cite{22} that combine FL with CL through generative replay, existing FCL works focus on solving the CL problem of multiple clients with similar data distributions, and using a global generator trained by FL to alleviate catastrophic forgetting, bringing three issues under high data heterogeneity: 
Firstly, the global generator is difficult to replay the local data distribution of a specific client, making it difficult to perform personalized aggregation. Secondly, the performance of the global generator will decrease as the data heterogeneity level increases. Finally, the global generator also needs to alleviate the catastrophic forgetting during its training through its own generated replay, thereby further reducing its own performance. 
Therefore, we need to redesign the generated replay architecture.

To address the above challenges in the FL setting of Data with Dynamic Heterogeneity under Limited Storage, we propose our pFL framework: pFedGRP. Due to the continuously arriving data over time in practice, it is difficult to determine whether the model has converged, we focus on the performance of the personalized global model on all known local data distributions in each FL round, rather than just its performance at the end of FL training.
Then we proposed a novel generative replay architecture: Firstly, due to the statistical heterogeneity of data mostly reflected in categories \cite{12}, we decouple the local generator of each client into multiple smaller sub models, each of which only performs updates on the real data of one category, thus there is almost no need to alleviate catastrophic forgetting. Secondly, we use local model to improve the generate performance of generator and to reduce the frequency of updating generator by detecting feature drift. Finally, to enhance the information of real data contained in the local model while mitigating catastrophic forgetting, we designed a local data distribution reconstruction scheme.
Based on the generated replay architecture above, we design a personalized aggregation scheme on server with learnable weights to flexibly trade-off the collaborative relationships between clients, and a local knowledge transfer scheme on client to improve the generalization and convergence rate of personalized global models. Our contribution is summarized as follows:

1. We extend the pFL to the FL setting of Data with Dynamic Heterogeneity under Limited Storage, then propose a novel optimization problem.

2. We propose a novel generative replay architecture that decouples the generator by category, improves generator performance through local models, and enhances the performance of local model by a local data distribution reconstruction scheme.

3. Based on the generated replay architecture above, we propose our pFL framework: pFedGRP, to conduct personalized aggregation and local knowledge transfer.

4. We conducted comparative experiments between pFedGRP and various FL, pFL, FCL methods on multiple benchmark datasets under various settings, the experimental results validated the effectiveness of our pFL framework.

\section{Related Work}

\subsection{Federated Learning and Personalized Federated Learning}
Federated Learning (FL) \cite{01} is a distributed machine learning paradigm without the transmission of dataset, the goal of FL is to aggregate a global model that performs well on all clients with different data heterogeneity.
One approach is improving the knowledge transfer within the model space. FedProx \cite{23} add a regularization term to restricting the $l_2$ distance between local models and global model parameters; FedLAW \cite{10} fine-tunes the aggregation weight on global validation dataset to improve generalization ability.
Another approach is to customize the personalized global models by adjusting the degree of collaboration between clients, which is denoted as personalized Federated Learning (pFL). FedEM \cite{25} regards the data distribution as a weighted mixture of multiple underlying data distributions, and uses EM algorithm to calculate the weight of underlying data distribution on the client side. pFedGraph \cite{26} calculates the cosine similarity of local models to construct a personalized collaboration graphs between clients. 
However, existing FL and pFL methods mostly contain the assumption of static local data distribution, which makes it difficult to cope with the changes of data heterogeneity within and between clients,  and cannot alleviate the catastrophic forgetting of models on inaccessible previous data.

\subsection{Federated Continue Learning}
Federated Continuous Learning (FCL) is an extension of Continuous Learning (CL) at the Federated Learning level where all clients have similar data distributions (i.e. the same task) at the same time, the goal of FCL is to keep the performance of the global model while the data distribution changes over time and the data of previous tasks cannot be accessed. 
One approach is directly combining FL with CL. FedWeIT \cite{27} decomposes the model into a weighted combination of global parameters for learning general knowledge and adaptive parameters for the task; FedET \cite{29} proposes a transformer based partial model component enhancement scheme.
Another approach is to use global knowledge to assist in local CL. TARGET \cite{22}, MFCL \cite{30} train a global generator with global model on the server to replay global features on the client; AF-FCL \cite{45} extracts global features by aggregating the local models and local generators obtained through alternating training on the client side.
Another way is to use model distillation to adjust the relationships between local knowledge. GLFC \cite{31} uses class aware gradient compensation and class semantic relation distillation to keep the consistency of the local inter-class relationships across different tasks; FedCIL \cite{20} uses ACGAN models to perform feature alignment and consistency enhancement with knowledge distillation during local training and global fine-tuning.
However, existing FCL methods contain the assumption that the local data distribution of different clients is similar at every moment, and typically assume that the change speed of data distribution (task) is slow to ensure model convergence, which makes it difficult to cope with the FL setting that the degree of data heterogeneity within and between clients changes over time.

\section{Preliminary}
In this section, we define the symbols in our paper, then elaborate on the optimization problem.
For the representation of the models, we use $C$ to represent the model used to solve practical problems (denoted as the Task Model), and use $A$ to represent the model used to generated replay (denoted as the Auxiliary Model).
For the representation of the distribution and the data, we use $\mathcal{P}=(\mathcal{X},\mathcal{Y})$ to represent the data distribution $\mathcal{P}$ as the joint distribution of the feature distributions $\mathcal{X}$ and the label distribution $\mathcal{Y}$, use $\left\{\mathcal{P}_1 \& \mathcal{P}_2\right\}$ and $\left\{\&_{i=1}^n \mathcal{P}_i\right\}$ to separately represent the weighted mixture of two distributions $\left\{\mathcal{P}_1, \mathcal{P}_2\right\}$ and $n$ distributions $\left\{\mathcal{P}_1,...,\mathcal{P}_n\right\}$ based on the data volume of each distribution, and use $\left\{\mathcal{D}_1\cup\mathcal{D}_2\right\}$ to represent the merging of two datasets $\left\{\mathcal{D}_1, \mathcal{D}_2\right\}$.

\subsection{Notations and Problem Formulation}
\textbf{Federated Learning and Personalized Federated Learning}: 
Assuming there are $n$ clients, the set of clients is $\bm{\mathcal{C}}=\{\mathcal{C}_1,\ldots,\mathcal{C}_n \}$. For each client $\mathcal{C}_i\in\bm{\mathcal{C}}$, we use $\mathcal{P}_{\mathcal{C}_i}=(\mathcal{X}_{\mathcal{C}_i}, \mathcal{Y}_{\mathcal{C}_i})$ to represent its local data distribution, and use $C_i$ and $C_{*,i}$ to separately represent the local task model and the global task model on client $\mathcal{C}_i$.
The Federated Learning (FL) aggregates the local task models $\{C_i\}_{i=1}^n$ to obtain a global task model $C_g$ that minimizes the expected value of the task driven loss $\mathcal{L}(\cdot,\cdot)$ on the local data distributions $\{\mathcal{P}_{\mathcal{C}_1},\ldots,\mathcal{P}_{\mathcal{C}_n}\}$ (i.e. $C_{*,i}=C_g$). 
The personalized Federated Learning (pFL) aggregates $n$ personalized global task models $\{C_{g,i}\}_{i=1}^n$ for each client $\mathcal{C}_i\in\bm{\mathcal{C}}$ (i.e. $C_{*,i}=C_{g,i}$). 
Therefore, the optimization objectives of FL and pFL can be summarized as follows:
\begin{equation}
   \label{F_1}
   \underset{C_{*,i}}{min}{\underset{(x,y)\sim\mathcal{P}_{\mathcal{C}_i}}{E}} 
   \left[ \mathcal{L}(C_{*,i},(x,y)) \right], \forall\mathcal{C}_i\in\bm{\mathcal{C}}
\end{equation}
However, existing FL and pFL methods mostly contain the assumption of static local data distribution, that is, for any FL round $t,t' \in \{1, ..., T\} $, it satisfies $\mathcal{P}_{\mathcal{C}_i}^t = \mathcal{P}_{\mathcal{C}_i}^{t'},\forall{\mathcal{C}_i\in\bm{\mathcal{C}}}$, which means that these methods can only improve the performance of the task models on the data distribution corresponding to the currently accessible data.

\textbf{Continual Learning and Federated Continual Learning}: 
Continuous Learning (CL) consists of a sequence $\bm{\mathcal{T}}=\{\mathcal{T}^1,\ldots,\mathcal{T}^T \}$ of $T$ tasks in time series. When executing the $t$-th task $\mathcal{T}^t \in \bm{\mathcal{T}}$, we denote the instant data distribution as $\mathcal{P}^t=(\mathcal{X}^t,\mathcal{Y}^t)$, the actual data distribution as $\left\{\&_{t'=1}^t \mathcal{P}^{t'}\right\}$, and it will not be possible to access the datasets of previous tasks $\{\mathcal{T}^1,\ldots,\mathcal{T}^{t-1} \}$. The goal of CL at each task $\mathcal{T}^t \in \bm{\mathcal{T}}$ is to obtain a task model $C^t$ that performs well in the actual data distribution $\left\{\&_{t'=1}^t \mathcal{P}^{t'}\right\}$.
Federated Continuous Learning (FCL) typically refers to the FL setting where all clients are in CL setting and have the same task in each FL round. Under this setting, clients execute every task of CL through multiple FL rounds together. Specifically, let task $\mathcal{T}^t$ consist of $R^t$ FL rounds, for each client $\mathcal{C}_i\in\bm{\mathcal{C}}$, the instant local data distribution $\mathcal{P}^r_{\mathcal{C}_i} = \mathcal{P}^t, \forall r \in \left\{1, \ldots, R^t \right\}$, and the actual local data distribution is still $\left\{\&_{t'=1}^t \mathcal{P}^{t'}\right\}$. Therefore, with all clients only can access to the dataset corresponding to $\mathcal{P}^t$, the optimization goal of FCL in each task $\mathcal{T}^t$ is to aggregate a global task model $C_g^t$ that performs well on the actual data distribution $\left\{\&_{t'=1}^t \mathcal{P}^{t'}\right\}$, that is:
\begin{equation}
   \label{F_2}
   \underset{C_{g}^t}{min}{\underset{(x,y)\sim\left\{\&_{t'=1}^t \mathcal{P}^{t'}\right\}}{E}} 
   \left[ \mathcal{L}(C_{g}^t,(x,y)) \right], 
   \forall \mathcal{T}^t \in \bm{\mathcal{T}}
\end{equation}
However, in reality, the instant local data distributions between client are usually different, and the time interval between the changes of data distribution may also be smaller than the FL rounds required by FCL methods to complete each task, making it difficult for these methods to achieve model convergence, thereby reducing model performance.

\textbf{Problem Formulation}: 
For simplicity, we consider the case where the instant local data distributions on the clients change with FL rounds. Due to the different data distributions of different clients, each client $\mathcal{C}_i$ has a CL task sequence $\bm{\mathcal{T}}_{\mathcal{C}_i}=\{\mathcal{T}^1_{\mathcal{C}_i},\ldots,\mathcal{T}^T_{\mathcal{C}_i} \}$ corresponding to $T$ FL rounds. 
At this point, client $\mathcal{C}_i$ executes task $\mathcal{T}^t_{\mathcal{C}_i}\in\bm{\mathcal{T}}_{\mathcal{C}_i}$ in each FL round $t\in\{1,\ldots,T\}$, the instant local data distribution and the actual data distribution of client $\mathcal{C}_i$ are $\mathcal{P}^t_{\mathcal{C}_i}$ and $\left\{\&_{t'=1}^t \mathcal{P}^{t'}_{\mathcal{C}_i}\right\}$, respectively, and client $\mathcal{C}_i$ cannot access the dataset of the previous $t-1$ FL rounds (tasks). 
The optimization objective of pFL in each FL round $t$ is extended to aggregate personalized global task models $\{C_{g,i}^t\}_{i=1}^n$ that perform well on the actual data distribution $\left\{\&_{t'=1}^t \mathcal{P}^{t'}_{\mathcal{C}_i}\right\}$ of each client $\mathcal{C}_i\in\bm{\mathcal{C}}$:
\begin{equation}
   \label{F_3}
   \begin{split}
      \left\{\underset{C_{g,i}^t}{min}{\underset{(x,y)\sim\left\{\&_{t'=1}^t \mathcal{P}^{t'}_{\mathcal{C}_i}\right\}}{E}} 
      \left[ \mathcal{L}(C_{g,i}^t) \right], 
      \forall\mathcal{C}_i\in\bm{\mathcal{C}}\right\}, \forall t\in [T]
   \end{split}
\end{equation}

\subsection{Optimization Problem}
The main challenges in solving optimization objective \hyperref[F_3]{3} are as follows: Firstly, due to the inability to access the dataset of previous FL rounds, the models on clients faces catastrophic forgetting in local training. Secondly, the data heterogeneity across clients will change with FL rounds, making it difficult for server to effectively adjust the collaboration between clients to achieve personalized aggregation. 
Inspired by the CL based on generated replay, we configure an auxiliary model $A_i$ for each client $\mathcal{C}_i$ to replay the previous feature distributions.
For the first challenge, we denote the instant local data distribution of client $\mathcal{C}_i$ in $t$-th FL round as $\mathcal{P}^t_{\mathcal{C}_i}=(\mathcal{X}^t_{\mathcal{C}_i}, \mathcal{Y}^t_{\mathcal{C}_i})$, and denote the auxiliary model updated through the previous $t-1$ FL rounds as $A_i^{t-1}$, the replayed feature distribution $\mathcal{X}^{t-1}_{A_i}$ of $A_i^{t-1}$ is close to the historical feature distribution $\left\{\&_{t'=1}^{t-1} \mathcal{X}^{t'}_{\mathcal{C}_i}\right\}$, making the replayed data distribution $\mathcal{P}^{t-1}_{A_i}=(\mathcal{X}^{t-1}_{A_i}, \&_{t'=1}^{t-1} \mathcal{Y}^{t'}_{\mathcal{C}_i})$ close to the historical feature distribution $\left\{\&_{t'=1}^{t-1} \mathcal{P}^{t'}_{\mathcal{C}_i}\right\}$. Therefore, client $\mathcal{C}_i$ can update the personalized global task model $C_{g,i}^{t-1}$ on the data distribution $\left\{\mathcal{P}^{t-1}_{A_i}\&\mathcal{P}^t_{\mathcal{C}_i}\right\}$ to alleviate catastrophic forgetting then get $C_{i}^{t, *}$, that is:
\begin{equation}
   \label{F_4}
   C_{i}^{t, *}\leftarrow\underset{C_{g,i}^{t-1}}{argmin}{\underset{(x,y)\sim\left\{\mathcal{P}^{t-1}_{A_i}\&\mathcal{P}^t_{\mathcal{C}_i}\right\}}{E}} 
   \left[ \mathcal{L}(C_{g,i}^{t-1}, (x,y)) \right]
\end{equation}
Afterwards, client $\mathcal{C}_i$ updates $A_i^{t-1}$ to $A_i^{t}$ to fit the actual feature distribution $\left\{\&_{t'=1}^{t} \mathcal{X}^{t'}_{\mathcal{C}_i}\right\}$. For the second challenge, we denote $\bm{W}_{i}^t=\{w_{i,1}^t,\ldots,w_{i,n}^t\}$ as the personalized aggregation weight of client $\mathcal{C}_i$ whose sum is $1$, and denote $\sum_{j=1}^{n}w_{i,j}^t C_{j}^{t,*}$ as the aggregated model. Since the replayed data distribution $\mathcal{P}^{t}_{A_i}=(\mathcal{X}^{t}_{A_i}, \&_{t'=1}^{t} \mathcal{Y}^{t'}_{\mathcal{C}_i})$ from $A_i^t$ is close to the actual local data distribution $\left\{\&_{t'=1}^{t} \mathcal{P}^{t'}_{\mathcal{C}_i}\right\}$, server can optimize $\bm{W}_{i}^t$ on $\mathcal{P}^{t}_{A_i}$ to obtain the optimal aggregation weight $\bm{W}_{i}^{t,*}=\{w_{i,1}^{t,*},\ldots,w_{i,n}^{t,*}\}$, that is:
\begin{equation}
   \label{F_5}
   \begin{split}
      \bm{W}_{i}^{t,*}\leftarrow\underset{\bm W_{i}^t}{argmin}
      {\underset{(x,y) \sim \mathcal{P}^{t}_{A_i}}{E}} 
      \left[\mathcal{L} \left(\sum_{j=1}^{n}w_{i,j}^t C_{j}^{t,*},(x,y)\right) \right], 
      s.t. \sum_{j=1}^{n}w_{i,j}^t = 1
   \end{split}
\end{equation}
Finally, client $\mathcal{C}_i$ obtains the personalized global task model $C_{g,i}^t\leftarrow\sum_{j=1}^{n}w_{i,j}^{t,*} C_{j}^{t,*}$ , and the $t$-th FL round ends.

However, there are still three challenges in efficiently solving optimization problems \hyperref[F_4]{4} and \hyperref[F_5]{5}: 
Firstly, the auxiliary model hard to fully fit the actual feature distribution \cite{32}. Especially, as the number of tasks increases, insufficient model parameters may lead to the underfitting of the feature distribution \cite{33}, ultimately reducing the effectiveness of local training and personalized aggregation \cite{34}\cite{35}. 
Secondly, even if the auxiliary model has sufficient parameters to fit the feature distribution, it still needs to use the generated replay of itself to alleviate its catastrophic forgetting on training, not only introducing more generated replay errors to itself, but also require longer training time and more computing resources.
Thirdly, existing CL and FCL methods with generated replay usually generate data of random category, when the local label distribution of the client is severely skewed, there will be a serious deviation between the replayed data distribution and the actual local data distribution.
Therefore, we need to redesign the generated replay architecture to address the three challenges above.

\section{Methodology}
\subsection{Generated Replay Architecture}

\textbf{Auxiliary model with category decoupling}: 
Since there is no existing generative model that simultaneously meets small model size, short training time, and good replay performance \cite{36}, it is inefficient to use a single auxiliary model to record the features of all types of data.
In machine learning, the statistical heterogeneity of data is mostly reflected in categories \cite{12}, so the data distribution $\mathcal{P}=(\mathcal{X}, \mathcal{Y})$ can be regarded as the weighted mixing of the feature distribution $\mathcal{X}_c$ through the appearing probability of each category $c\in\mathcal{Y}$.
Since the number of categories in data is usually much smaller than the number of data, for each category $c\in\mathcal{Y}_{\mathcal{C}_i}$ encountered by client $\mathcal{C}_i$, we use a smaller auxiliary sub model $A_{i,c}$ to fit the $\mathcal{X}_{\mathcal{C}_i,c} \in \mathcal{X}_{\mathcal{C}_i}$ (i.e. $A_{i}=\left\{A_{i,c}\right\}_{c\in\mathcal{Y}_{\mathcal{C}_i}}$). 
Rather than updated on all currently accessible real data, $A_{i,c}$ only performs updates when the real data of category $c$ is accessible, making it almost unnecessary to consider catastrophic forgetting, thereby accelerating model training and reducing computation and communication cost.
However, there are still two issues: Firstly, the $A_{i,c}$ with smaller model size may be hard to fully fit $\mathcal{X}_{\mathcal{C}_i,c}$, thereby reducing the performance of generated replay. Secondly, if there is no feature drift between the real data of category $c$ in multiple FL rounds, updating $A_{i,c}$ will hardly bring any benefits.

\textbf{Improving performance through task model}: 
In local training, since the local task model $C_i$ performs update before updating $A_i$, we use the latest information of $\mathcal{X}_{\mathcal{C}_i,c}$ contained in $C_i^*$ which updated from $C_i$ to solve the issues above. 
For the first point, denoting $\mathcal{D}_{A_{i,c}}$ as the dataset generated by $A_{i,c}$, let $\mathcal{D}_{A_{i,c}, C_i^*}$ consists of the data in $\mathcal{D}_{A_{i,c}}$ that judged as category $c$ by $C_i^*$, server optimizes $\bm W_{i}$ on $\mathcal{D}_{A_{i,c}, C_i^*}$ to aggregate the personalized global task model $C_{g,i}$ for client $\mathcal{C}_i$. Then, let $\mathcal{D}_{A_{i,c}, C_{g,i}}$ consists of the data in $\mathcal{D}_{A_{i,c}}$ that judged as category $c$ by $C_{g,i}$, client $\mathcal{C}_i$ alleviates the catastrophic forgetting of $C_i$ on $\mathcal{D}_{A_{i,c}, C_{g,i}}$.
For the second point, when encountering real data of category $c$, client $\mathcal{C}_i$ calculates the proportion of the data in $\mathcal{D}_{A_{i,c}}$ that judged as category $c$ by $C_i^*$. The operation of updating $A_{i,c}$ only occurs when the proportion is below a certain threshold.
The premise of the above is to reduce the fitting error of $C_i^*$ on real data, that is, to increase the proportion of the real data on the training data of each category $c\in\mathcal{Y}_{\mathcal{C}_i}^{t}$ in each FL round $t\in\{1, \ldots, T\}$.

\textbf{Local Data Distribution Reconstruction Scheme}: To improve the proportion of real data on each category $c\in\mathcal{Y}_{\mathcal{C}_i}^{t}$ while approaching the actual local data distribution $\left\{\&_{t'=1}^{t} \mathcal{P}^{t'}_{\mathcal{C}_i}\right\}$, we propose the following scheme:
In $t$-th FL round, denoting $Y_{\mathcal{C}_i}^t$ as the vector composed of the number of each type of real data, client $\mathcal{C}_i$ calculates the real data volume vector $\sum_{t'=1}^{t}Y_{\mathcal{C}_i}^{t'}$ of all $t$ FL rounds, then proportionally shrinks it to a quantity where only one type of real data exists which is equal to the number of that type of data in $Y_{\mathcal{C}_i}^t$, denoted as $\left(\sum_{t'=1}^{t}Y_{\mathcal{C}_i}^{t'}\right)_s$, obtaining the maximum proportion of real data while the label distribution is equal to $\left\{\&_{t'=1}^{t} \mathcal{Y}^{t'}_{\mathcal{C}_i}\right\}$.
To reduce replay errors, we limit the volume of each type of generated data to no more than the volume of the type of real data with the highest volume in $Y_{\mathcal{C}_i}^t$, denoted as $\left(\sum_{t'=1}^{t}Y_{\mathcal{C}_i}^{t'}\right)_{ss}$. Finally, client $\mathcal{C}_i$ calculates the generated data volume vector $Y_{\mathcal{C}_i, A}^{t}=\left(\sum_{t'=1}^{t}Y_{\mathcal{C}_i}^{t'}\right)_{ss}-Y_{\mathcal{C}_i}^t$, the flowchart is shown in Figure \hyperref[Figure 2]{2}:
\begin{figure}[ht]
   \vskip 0.2in
   \label{Figure 2}
   \begin{center}
   \centerline{\includegraphics[width=0.8\columnwidth]{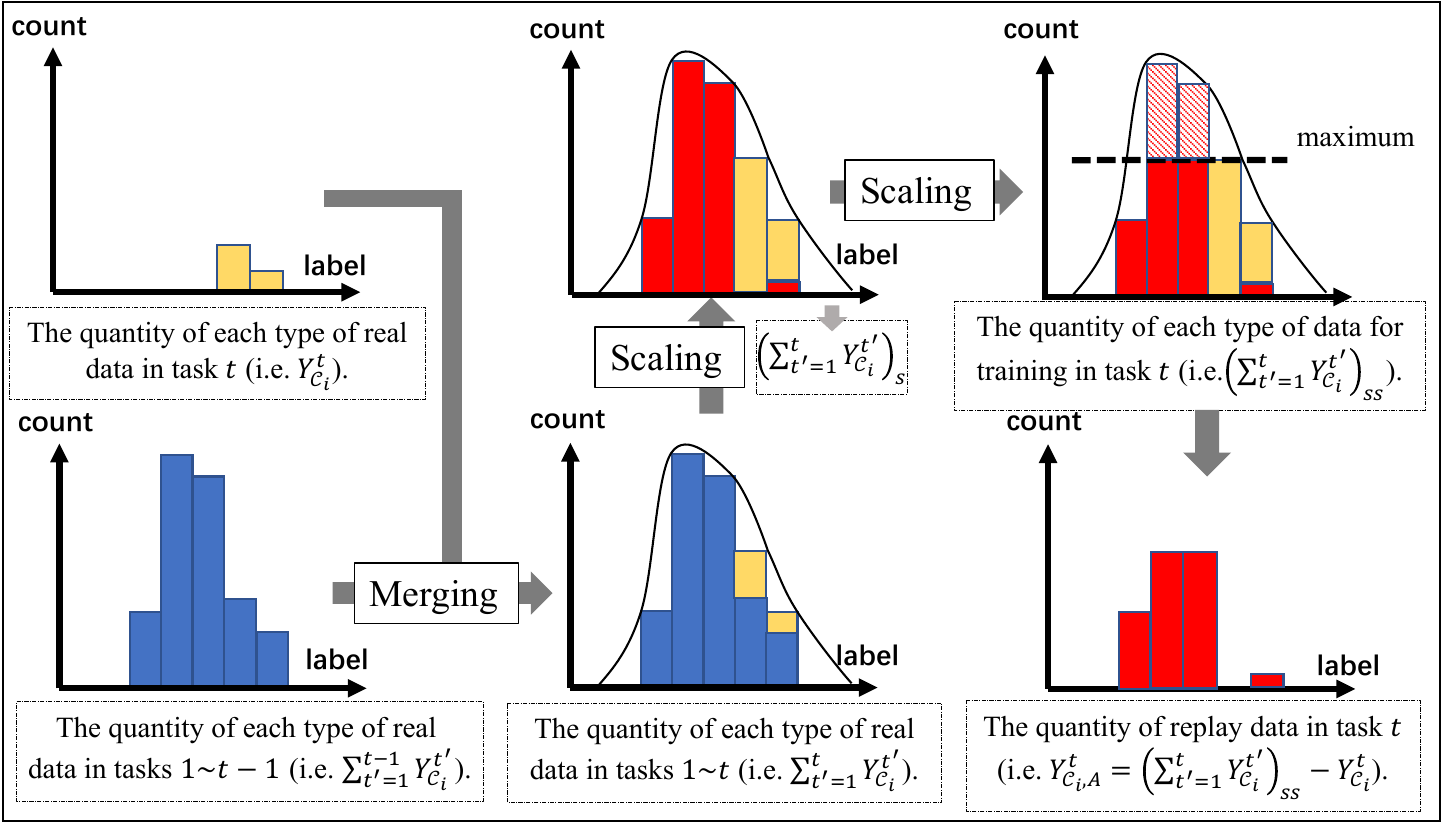}}
   \caption{\text{\small\sl Figure 2.} Local data distribution reconstruction scheme.}
   \end{center}
   \vskip -0.2in
\end{figure}

\subsection{pFedGRP}
With the Generated Replay Architecture above, we propose our pFL framework: pFedGRP, and take the $t\in\{1,\ldots,T\}$ FL round to illustrate its process.

\textbf{Local Training}:
Before local training, client $\mathcal{C}_i\in\bm{\mathcal{C}}$ has three models: auxiliary model $A_i^{t-1}$, personalized global task model $C_{g,i}^{t-1}$, and a global task model $C_{g}^{t-1}$ obtained by average aggregation. 
Firstly, client $\mathcal{C}_i$ calculates the generated data volume vector $Y_{\mathcal{C}_i, A}^{t}$ through Local Data Distribution Reconstruction Scheme, then uses $A_i^{t-1}$ and $C_{g,i}^{t-1}$ to create the generate replay dataset $\mathcal{D}_{A_{i},C_{g,i}}^{t-1}$ through $Y_{\mathcal{C}_i, A}^{t}$, and mixes it with real dataset $\mathcal{D}_{\mathcal{C}_i}^{t}\sim\mathcal{P}_{\mathcal{C}_i}^{t}$ to form the training dataset $\left\{\mathcal{D}_{A_{i},C_{g,i}}^{t-1}\cup\mathcal{P}_{\mathcal{C}_i}^{t}\right\}$ of the local task model.
Under high data heterogeneity, to improve the generalization of the local task model, client $\mathcal{C}_i$ performs local training on the global task model $C_{g}^{t-1}$, and aligns the outputs of $C_{g}^{t-1}$ and $C_{g,i}^{t-1}$ on $\mathcal{D}_{A_{i},C_{g,i}}^{t-1}$ through mean square error (MSE) to reduce feature drift, with the weight denoted as $\lambda$, that is:
\begin{equation}
   \label{F_6}
   \begin{split}
      C_{i}^{t,*}\leftarrow\underset{C_{g}^{t-1}}{argmin}
      \left\{
      \sum_{(x,y)\in \left\{\mathcal{D}_{A_{i},C_{g,i}}^{t-1}\cup\mathcal{D}_{\mathcal{C}_i}^{t} \right\}}
      \mathcal{L} \left(C_{g}^{t-1},(x,y)\right) \right.
      \left. \quad + \quad \lambda\cdot\sum_{x \in \mathcal{D}_{A_{i},C_{g,i}}^{t-1}}
      MSE \left(C_{g}^{t-1}(x),C_{g,i}^{t-1}(x)\right)
      \right\}
   \end{split}
\end{equation}
Afterwards, client $\mathcal{C}_i$ uses $C_{i}^{t,*}$ to judge whether each sub model in $A_i^{t-1}$ needs to be updated. If the auxiliary sub model $A_{i,c}^{t-1}$ needs to be updated, denote the loss as $\mathcal{L}_A$ and the real data subset as $\mathcal{D}_{\mathcal{C}_i, y=c}^{t}\subset\mathcal{D}_{\mathcal{C}_i}^{t}$, that is:
\begin{equation}
   \label{F_7}
   \begin{split}
      A_{i,c}^{t,*}\leftarrow&\underset{A_{i,c}^{t-1}}{argmin}
      \sum_{x \in \mathcal{D}_{\mathcal{C}_i, y=c}^{t}}
      \mathcal{L}_A \left(A_{i,c}^{t-1},x\right)
   \end{split}
\end{equation}
Denoting the set of all updated auxiliary sub models as $\left\{A_{i,c}^{t,*}\right\}$, client $\mathcal{C}_i$ uses $\left\{A_{i,c}^{t,*}\right\}$ to update $\left\{A_{i,c}^{t-1}\right\}$ then get  $\left\{A_{i,c}^{t}\right\}$, and uses $\left\{A_{i,c'}^{t-1}\right\}$ as $\left\{A_{i,c'}^{t}\right\}$ for other category $c'$, thus, updating the auxiliary model $A_i^{t-1}$ to $A_i^t$. After local training, client $\mathcal{C}_i$ sends $C_{i}^{t,*}$, $\left\{A_{i,c}^{t,*}\right\}$ and the actual local label distribution $\left\{\&_{t'=1}^{t} \mathcal{Y}^{t'}_{\mathcal{C}_i}\right\}$ to the server.
The flowchart of local training is Figure \hyperref[Figure 3]{3}:
\begin{figure}[ht]
   \vskip 0.2in
   \begin{center}
   \label{Figure 3}
   \centerline{\includegraphics[width=\columnwidth]{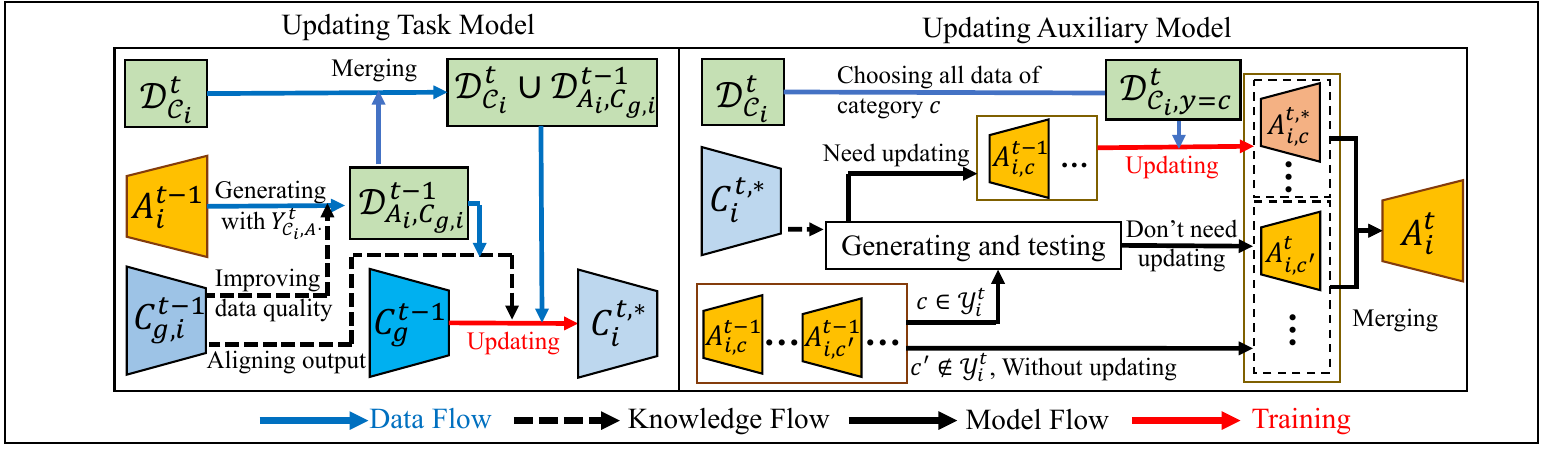}}
   \caption{\text{\small\sl Figure 3.} The flowchart of Local Training on client $\mathcal{C}_i$.}
   \end{center}
   \vskip -0.2in
\end{figure}

\begin{figure}[ht]
  \vskip 0.2in
  \begin{center}
  \label{Figure 4}
  \centerline{\includegraphics[width=\columnwidth]{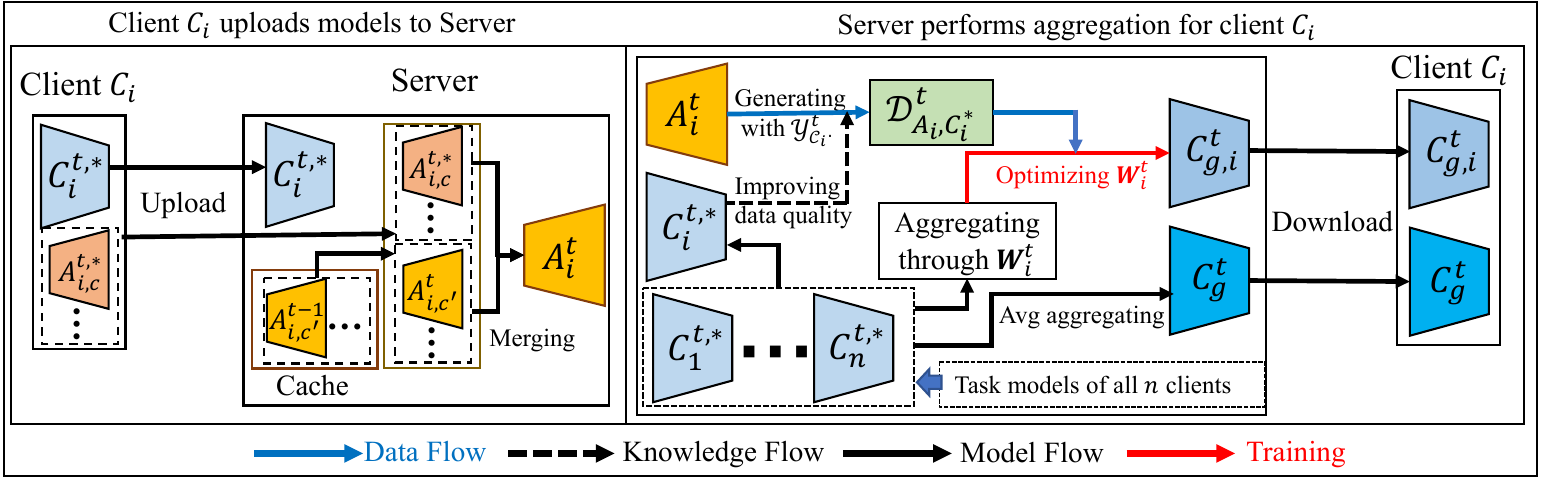}}
  \caption{\text{\small\sl Figure 4.} The flowchart of Global Aggregation on server.}
  \end{center}
  \vskip -0.2in
\end{figure}

\textbf{Personalized Aggregation}: 
The server receives the data sent by all $n$ clients and denotes the set of local task models as $\left\{C_{1}^{t,*}, \ldots, C_{n}^{t,*}\right\}$. For each client $\mathcal{C}_i\in\bm{\mathcal{C}}$, the server updates the auxiliary model cache $A_i^{t-1}$ with $\left\{A_{i,c}^{t,*}\right\}$ to synchronize $A_i^{t}$, then uses $A_i^{t}$ and $C_{i}^{t,*}$ to create the generate replay dataset $\mathcal{D}_{A_i, C_{i}^*}^t$ through $\left\{\&_{t'=1}^{t} \mathcal{Y}^{t'}_{\mathcal{C}_i}\right\}$. Afterwards, the server optimizes the personalized aggregation weight $\bm{W}_{i}^{t}=\{w_{i,1}^{t},\ldots,w_{i,n}^{t}\}$ on $\mathcal{D}_{A_i, C_{i}^*}^t$ to obtain $\bm{W}_{i}^{t,*}$, that is:
\begin{equation}
   \label{F_8}
   \begin{split}
      \bm{W}_{i}^{t,*}\leftarrow\underset{\bm{W}_{i}^{t}}{argmin}
      \sum_{(x,y) \in \mathcal{D}_{A_i, C_{i}^*}^t}
      \mathcal{L} \left(\sum_{j=1}^{n}(w_{i,j}^{t}C_{j}^{t,*}),(x,y)\right), 
      s.t. \sum_{j=1}^{n}w_{i,j}^t = 1
   \end{split}
\end{equation}
Finally, the server aggregates a personalized global task model $C_{g,i}^t\leftarrow\sum_{j=1}^{n}(w_{i,j}^{t,*}C_{j}^{t,*})$ for client $\mathcal{C}_i$. After completing the personalized aggregation of all clients, the server averaged aggregates a global task model $C_{g}^t \leftarrow \frac{1}{n} \sum_{j=1}^{n}C_{j}^{t,*}$, and then sends $C_{g,i}^t$ and $C_{g}^t$ to each client $\mathcal{C}_i\in\bm{\mathcal{C}}$. 
The flowchart of global aggregation is Figure \hyperref[Figure 4]{4}.

The pseudocode of pFedGRP can be found in Appendix \hyperref[Appendix C.1]{C.1}.

\section{Experiment}

\subsection{Datasets and Settings}
\label{Datasets and Settings}
We construct the FL setting of Data with Dynamic Heterogeneity under Limited Storage based on existing MNIST dataset \cite{37}, FashionMNIST dataset \cite{38}, Cifar10 dataset \cite{N39}, Cifar100 dataset \cite{N39} and EMNIST ByClass dataset \cite{40}: For all datasets, we set the total number of clients to 10, each client randomly divides the set of data categories of the dataset into multiple subsets, each subset containing two categories and corresponding to a type of tasks. Due to the different partitioning results of the data categories between clients, the types of tasks contained in different clients are likely to be different. 
Specifically, each client randomly divides the 10 categories of the MNIST, FashionMNIST and Cifar10 datasets into 5 types of tasks, the 62 categories of the EMNIST-ByClass dataset into 31 types of tasks, the 100 categories of the Cifar100 dataset into 50 types of tasks.
In each FL round, each client selects a type of task to execute, the accessible data of the client in this FL round consists of the real data of two categories corresponding to the task, and the number of real data in each category is 200, the total is 400. 
Unlike other FCL methods that switch the type of task between multiple FL rounds, we switch the type of task between every two FL rounds to better simulate our FL setting.
For each client, each training data in the dataset can only be accessed in one FL round, and cannot be accessed in any subsequent FL round, but the test data of that FL round will be used for testing in subsequent FL rounds. We provide detailed information of the datasets and settings in Appendix \hyperref[Appendix A]{A}.

Based on the data complexity of the dataset, We select different generative models as auxiliary sub model for pFedGRP: For the MNIST series dataset, we choose the 16 channels WGAN-GP \cite{42} model whose network structure is similar to DCGAN \cite{41}, denoted as pFedGRP+WGAN-GP. For the Cifar series dataset, we choose two auxiliary sub models: 1. The above WGAN-GP model with 64 channels, denoted as pFedGRP+WGAN-GP. 2. the DDPM \cite{44} model sampled with DPM solver \cite{43}, denoted as pFedGRP+DDPM. We provide the floating point operations (FLOPs) and parameter count of the auxiliary sub models above in Appendix \hyperref[Appendix C.4]{C.4}.

\subsection{Baselines and Metrics}
We compare pFedGRP with various FL, pFL and FCL baseline methods. FL methods include two classic methods, FedAVG \cite{01} and FedProx \cite{23}; pFL methods include a classic FedEM \cite{25} and a newer pFedGraph \cite{26}; FCL methods include four methods: FedCIL \cite{20}, TARGET \cite{22}, MFCL \cite{30}, AF-FCL \cite{45}.
We set the performance where clients can access the real data of all previous FL rounds as the upper bound, denoted as "Centralized". 
We provide detailed information of these baseline methods in Appendix \hyperref[Appendix B]{B}, and provide FLOPs and parameter count of these FCL methods in Appendix \hyperref[Appendix C.4]{C.4}.

For evaluation metrics, we define Instant Average Accuracy (IAA) to measure the performance of each method in each FL round, and calculate the Average Accuracy (AA) of each method to measure the absolute performance. Meanwhile, we use the mean difference between the IAA of the centralized method and the IAA of other methods as the average forgetting metric (AFM) to measure the forgetting degree of each method. We provide details of the metrics in Appendix \hyperref[Appendix C.2]{C.2}.

\subsection{Baseline Experiments}
We designed experiments to compare pFedGRP with other baseline FL methods in three scenarios. The first two scenarios are conducted on the MNIST, FashionMNIST, and Cifar10 datasets, the last scenario is conducted on the EMNIST-ByClass and Cifar100 datasets. Since that the clients are unable to access the real data encountered in the previous FL round, on the MNIST and FashionMNIST datasets, each client can build up to 150 tasks for 150 FL rounds in five types of tasks with non-overlapping real data; on the Cifar10 dataset, each client can build up to 125 tasks for 120 FL rounds in five types of tasks with non-overlapping real data.

\begin{figure}[ht]
   \vskip -0.1in
   \caption{\text{\small\sl Table 1.} Results on FL with Tasks Gradually Changing.}
   \label{Table 1}
   \vskip 0.15in
   \begin{center}
   \centerline{\includegraphics[width=0.6\columnwidth]{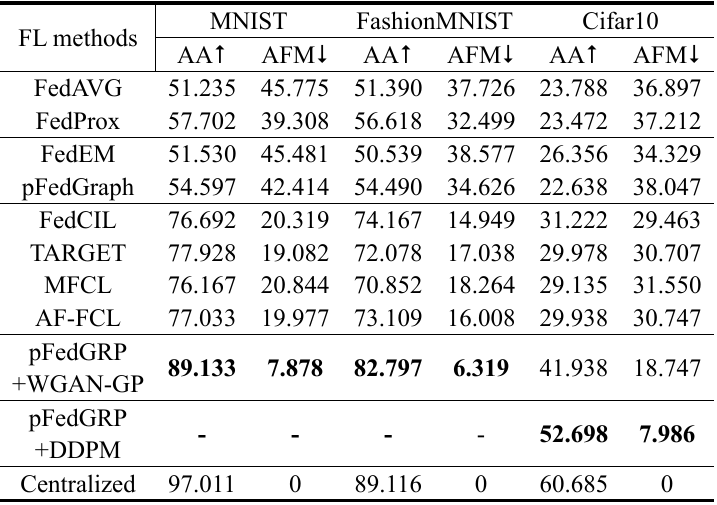}}
   \end{center}
   \vskip -0.2in
\end{figure}

\textbf{FL with Tasks Gradually Changing}: In this setting, each client $\mathcal{C}_i$ randomly selects two types of tasks from its five types of tasks (denoted as $T_{\mathcal{C}_i, 1}$, $T_{\mathcal{C}_i, 2}$) to form a task loop, that is, as the FL rounds increase, the client $\mathcal{C}_i$ executes $T_{\mathcal{C}_i, 1}$, $T_{\mathcal{C}_i, 2}$, $T_{\mathcal{C}_i, 1}$, $T_{\mathcal{C}_i, 2}$……After 30 FL rounds on MNIST and FashionMNIST (24 FL rounds on Cifar10), client randomly selects another type of task (denoted as $T_{\mathcal{C}_i, 3}$) to replace one type of task in the task loop. Specifically, if $T_{\mathcal{C}_i, 1}$ is replaced, the task loop consists of $T_{\mathcal{C}_i, 2}$ and $T_{\mathcal{C}_i, 3}$. This setting corresponds to the common situation where the data distribution changes slowly in real-time. The experimental results are shown in Table \hyperref[Table 1]{1}.

Before the task model converges, our pFedGRP uses personalized aggregation to better maintain the performance of the task model on all categories of data encountered previously, thereby achieving better overall performance while reducing forgetting. The IAA variation and analysis are shown in Appendix \hyperref[Appendix E.1]{E.1}, and the calculation and communication consumption are shown in Appendix \hyperref[Appendix C.4]{C.4}.

\begin{figure}[ht]
   \vskip -0.1in
   \caption{\text{\small\sl Table 2.} Results on FL with Tasks Circulating.}
   \label{Table 2}
   \vskip 0.15in
   \begin{center}
   \centerline{\includegraphics[width=0.6\columnwidth]{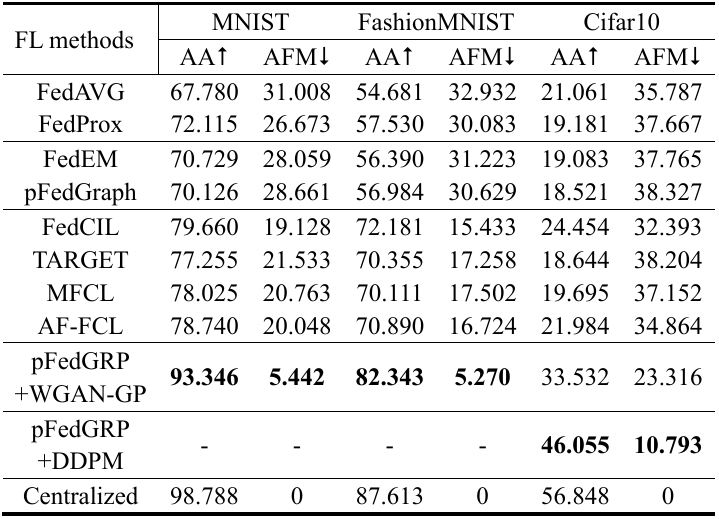}}
   \end{center}
   \vskip -0.2in
\end{figure}

\textbf{FL with Tasks Circulating}: In this setting, each client $\mathcal{C}_i$ forms its five types of tasks into a task cycle in random order, that is, as the FL rounds increased, the client $\mathcal{C}_i$ executed $T_{\mathcal{C}_i, 1}$, $T_{\mathcal{C}_i, 2}$, $T_{\mathcal{C}_i, 3}$, $T_{\mathcal{C}_i, 4}$, $T_{\mathcal{C}_i, 5}$, $T_{\mathcal{C}_i, 1}$……This setting corresponds to the situation where the data distribution changes extremely drastic. The experimental results are shown in Table \hyperref[Table 2]{2}.

The IAA variation and analysis are shown in Appendix \hyperref[Appendix E.2]{E.2}, and the calculation and communication consumption are shown in Appendix \hyperref[Appendix C.4]{C.4}.

\begin{figure}[ht]
   \vskip -0.1in
   \caption{\text{\small\sl Table 3.} Results on FL under High Data Heterogeneity.}
   \label{Table 3}
   \vskip 0.15in
   \begin{center}
   \centerline{\includegraphics[width=0.6\columnwidth]{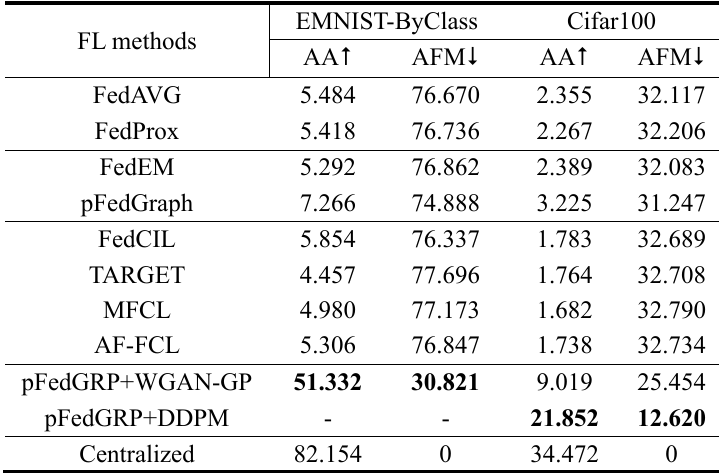}}
   \end{center}
   \vskip -0.2in
\end{figure}

\textbf{FL under High Data Heterogeneity}: We also compared the performance of the FL methods above under high data heterogeneity settings on the Cifar100 dataset and the EMNIST ByClass dataset:
Each client $\mathcal{C}_i$ forms its all types of tasks (50 for Cifar100, 31 for EMNIST-ByClass) into a task cycle in random order, then complete one task cycle. At this point, all FL methods cannot reach convergence, which better reflects the robustness of these FL methods. The experimental results are shown in Table \hyperref[Table 3]{3}.

It shows that pFedGRP has stronger robustness than other FL methods. The IAA variation and analysis are shown in Appendix \hyperref[Appendix E.3]{E.3}, and the calculation and communication consumption are shown in Appendix \hyperref[Appendix C.4]{C.4}.

\textbf{More Experiments}: We conduct ablation experiments of pFedGRP, the experimental details and results can be found in Appendix \hyperref[Appendix D.1]{D.1}. With similar settings as FL with Tasks Gradually Changing, we increase the correlation between tasks to explored the performance changes of various FL methods, the experimental details and results can be found in Appendix \hyperref[Appendix D.2]{D.2}.

\section{Conclusion}

In this work, we extend the personalized federated learning to the FL setting of Data with Dynamic Heterogeneity under Limited Storage, and attempt to solve this problem through generated replay. We first propose a novel generative replay architecture that alleviates the catastrophic forgetting of the auxiliary models by decoupling them by category, improves the generation performance of auxiliary models and reduces their update frequency through task models, and improves the performance of local task models by reconstructing local data distributions. Based on the generated replay architecture above, we propose a personalized aggregation scheme and a local knowledge transfer scheme. The above constitute our pFedGRP framework. We validated the performance of pFedGRP in experiments with various  datasets and settings.

\bibliographystyle{unsrt}  
\bibliography{references}  

\newpage

\appendix
\section{Datasets and Setting}
\label{Appendix A}
On the FL setting of Data with Dynamic Heterogeneity under Limited Storage, we use existing datasets to build the local dataset for each client. In our setting, the time interval between the server sends the global task model to the clients is one FL round, each client executes a specific task within its types of tasks (see section \hyperref[Datasets and Settings]{5.1}) in each FL round. Specifically, each type of task contains multiple specific tasks with the same category but non duplicate data, and each task contains training data and testing data, the training data can only be accessed by the client during the FL round of executing this specific task, but the test data will be used for all FL rounds after executing this specific task to test the performance of the task model on the client side.

\begin{figure*}[h]
  \vskip 0.2in
  \begin{center}
  \label{Figure 5}
  \centerline{\includegraphics[width=\columnwidth]{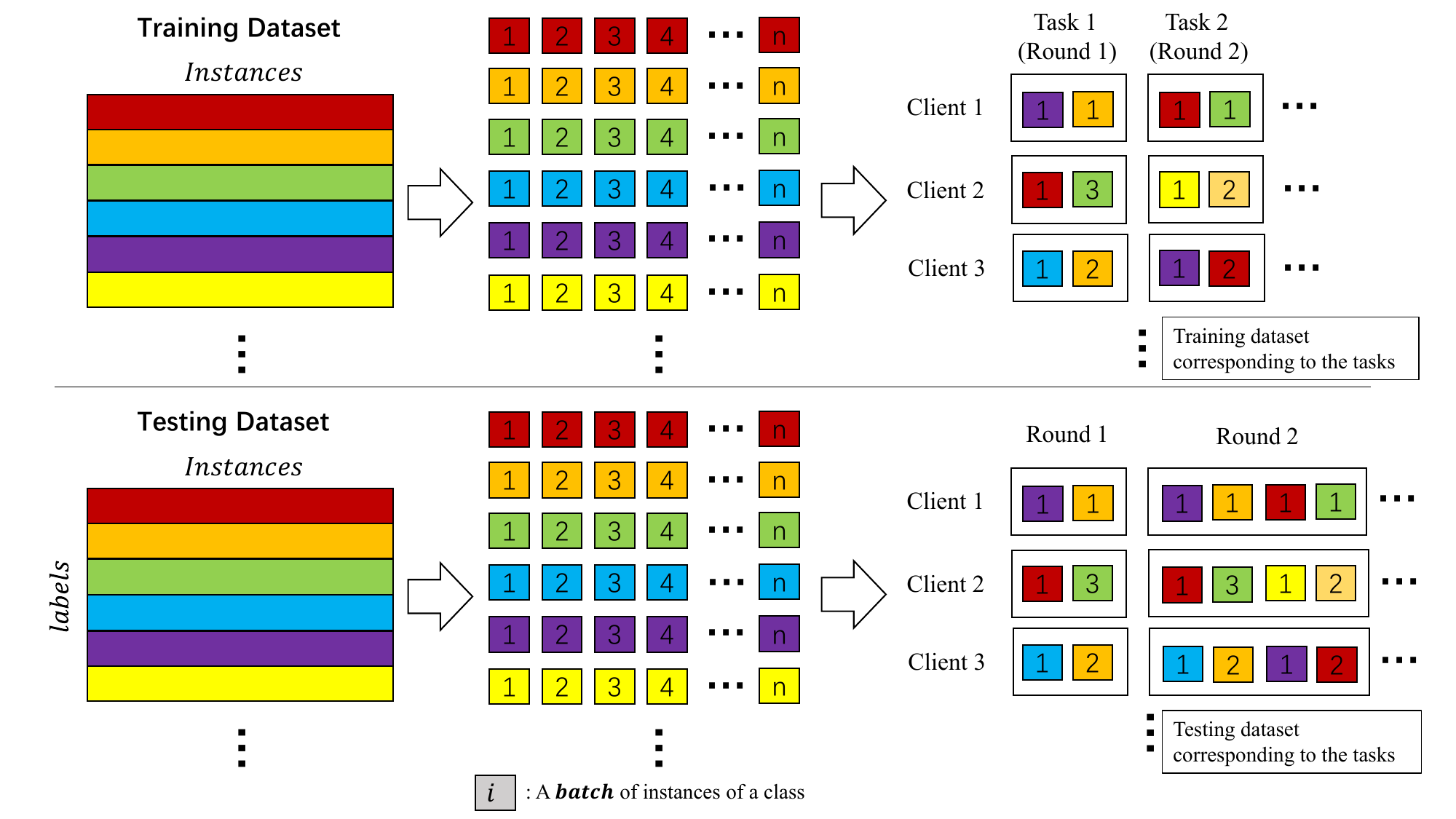}}
  \caption{\text{\small\sl Figure 5.} Schematic diagram of the partitioning of local training data and testing data.}
  \end{center}
  \vskip -0.2in
\end{figure*}

The schematic diagram of the partitioning of local training data and testing data are shown in Figure \hyperref[Figure 5]{5}: Each color in Figure \hyperref[Figure 5]{5} represents a type of data, we split each type of data on the training and testing sets into $n$ non-overlapping parts in groups of 200 data. In each FL round, based on the data categories corresponding to the tasks, each client selects training data parts that have not been accessed by it to build the training dataset (as shown in the upper right side of the figure), and add the corresponding test data parts into the test dataset (as shown in the lower right side of the figure).

The specific information of each dataset we used for the experiment is as follows:

\textbf{MNIST}. The MNIST dataset \cite{37} is a 10 categories numerical classification dataset with 60000 training samples and 10000 test samples, and each sample is a single channel grayscale image with a size of 28x28 containing a number from 0 to 9. In our baseline experimental setup, the total number of clients is 10, each client contains 5 tasks, each task consists of 2 random and non repeating types of data with 200 data in each type. 

\textbf{FashionMNIST}. The FashionMNIST dataset \cite{38} is a clothing classification dataset consisting of 10 categories, each category with 6000 training samples and 1000 testing samples, and all samples are single channel grayscale images with a size of 28x28. Compared to the MNIST dataset, FashionMNIST dataset includes projections of objects from different perspectives which making it more challenging in terms of image quality and diversity. Our experimental setup on the FashionMNIST dataset is the same as that on the MNIST dataset.

\textbf{EMNIST-ByClass}. The EMNIST-ByClass dataset \cite{40} is a dataset consisting of 62 imbalanced categories of handwritten characters and numbers with 814255 grayscale images of size 28x28. Compared with the MNIST dataset, EMNIST-ByClass dataset contains more categories, and its English character part includes uppercase and lowercase characters which increases the difficulty of classification. We strictly adhere to the definition of federated class incremental learning on this dataset: The total number of clients is 10, each client contains 31 tasks consisting of randomly non repeating two types of data with 200 training data and 100 testing data for each type.

\textbf{CIFAR10}. The CIFAR10 dataset \cite{N39} is a real image classification dataset consisting of 10 categories of 32x32 color RGB images, each category containing 5000 training images and 1000 test images. Compared with the MNIST series dataset, CIFAR-10 contains objects in the real world which have not only have a lot of noise but also different proportions and features, making data classification more difficult. Our experimental setup on the CIFAR10 dataset is the same as that on the MNIST dataset. 

\textbf{CIFAR100}. The CIFAR100 dataset \cite{N39} is a real image classification dataset consisting of 20 super categories, each super category has 5 categories and contains of 32x32 color RGB images. Each category contains 500 training images and 100 test images. Compared with the CIFAR10 dataset, the CIFAR100 dataset has a larger number of categories, and the images of each category within the same super category are more similar which increases the difficulty of classification. We strictly adhere to the definition of federated class incremental learning on this dataset: The total number of clients is 10, each client contains 50 tasks consisting of randomly non repeating two types of data with 200 training data and 100 testing data for each type.

\section{Baselines Details}
\label{Appendix B}
We compare our personalized federated learning framework pFedGRP with following two FL methods, two pFL methods and four FCL methods. The FL methods and pFL methods do not have the ability to remember information related to historical tasks while the FCL methods can solve catastrophic forgetting and statistical heterogeneity problems. We additionally incorporated FL and pFL methods combined with our generative replay framework in the ablation experiment to validate the effectiveness of the personalized aggregation scheme of pFedGRP. 

\textbf{FedAVG}: FedAVG \cite{01} is a representative federated learning method. Based on the size of the client's local training dataset, server weighted aggregates the local task models uploaded by clients to obtain a global task model.

\textbf{FedProx}: FedProx \cite{23} made some improvements to FedAVG, adding a proximal term to the local training loss to avoid the local task model deviating too much from the global task model. The aggregation strategy of FedProx is consistent with FedAVG.

\textbf{FedEM}: FedEM \cite{25} is a classic personalized federated learning method, it proposed that the local data distribution is a weighted mixture of several underlying data distributions. Correspondingly, it trains several sub task models on each client to fit these underlying distributions, and aggregate each sub model separately. Then, the client performs EM steps on the local dataset based on several global task sub models aggregated by the server through FedAVG's strategy to calculate the personalized weights of each sub model. Finally, clients calculate personalized weights by performing EM steps on global task sub models on their local dataset.

\textbf{pFedGraph}: pFedGraph \cite{26} is a relatively new personalized federated learning method, it proposes to use the cosine degree of the local task models to solve a personalized collaboration graph on server, then provides personalized aggregation for each client to balance the relationship between individual utility and collaboration benefit. In addition, it uses the cosine similarity of model parameters to constrain the bias of local task model in local training.

\textbf{FedCIL}: FedCIL \cite{20} is a relatively new federated class incremental learning method which integrates the task model and auxiliary model into one ACGAN model. In the local training phase, with the generated data of the global ACGAN model and the previous local ACGAN model, FedCIL uses model distillation and label alignment to alleviate the catastrophic forgetting of the local ACGAN model. In the global aggregation phase, server first averaged aggregates the local ACGAN models to obtain a global ACGAN model, then finetune the global ACGAN model with the generated data of each local ACGAN model.

\textbf{TARGET}: TARGET \cite{22} is a relatively new federated class incremental learning method based on global feature replay. On the server side, it trains a global generator with the BN layer features of the aggregated global task model and an untrained task model. On the client side, it alleviates the catastrophic forgetting of the task model with the data replayed by the global generator.

\textbf{MFCL}: MFCL \cite{30} is a relatively new federated class incremental learning method based on global sample free replay and distillation. On the server side, it proposed a scheme to training a global generator capable of generating high-quality data with the global task model aggregated. On the client side, it transfers the knowledge of the global task model to the local task model through distillation with the generated data of the global generator

\textbf{AF-FCL}: AF-FCL \cite{45} is a relatively new federated class incremental learning method based on local sample free replay. Based on the idea of partial feature forgetting, it designs a local distillation mechanism. On the client side, to achieve the goals of extracting data features for local task model and obtain an auxiliary model with better replay effects, it trains the local task model and the local auxiliary model alternately with the real data and the data generated by global auxiliary model. On the server side, it uses average aggregation to aggregate task models and auxiliary models.

\textbf{FedAVG-replay}: The FedAVG algorithm that additionally uses the generate replay scheme of pFedGRP on local training.

\textbf{pFedGraph-replay}: The pFedGraph algorithm that additionally uses the generate replay scheme of pFedGRP on local training.

\textbf{Centralized}: During local training, client can access the real data encountered in previous FL rounds. After local training, server does not aggregate local task models to create a global task model.

\section{Implementation Details}
\label{Appendix C}

\subsection{Algorithm and flowchart of pFedGRP}
\label{Appendix C.1}

The algorithm for pFedGRP is as follows:

\begin{algorithm}[h]
   \caption{pFedGRP}
   \label{Algorithm pFedGRP}
   \begin{algorithmic}
      \STATE {\bfseries Input:} Client set $\bm{\mathcal{C}}=\left\{\mathcal{C}_1,\ldots,\mathcal{C}_n\right\}$ with $n$ clients; Task model $C$ and auxiliary sub models $A=\left\{A_1,A_2,\ldots\right\}$.
      \STATE {\bfseries Output:} Personalized global task models $\left\{C_{g,1}^t, \ldots,C_{g,n}^t\right\}$ of $n$ clients in each FL round $t\in\left\{1,\ldots,T\right\}$.

      \STATE Server random initializes $C$ and takes it as global task model $C_g^0$ and personalized global task models $\left\{C_{g,1}^0,\ldots,C_{g,n}^0\right\}$.

      \FOR{each FL round $t=1,\ldots,T$}

         \STATE // Client local training

         \FOR{each client $\mathcal{C}_i\in\bm{\mathcal{C}}$ in parallel}

            \STATE server sends $C_{g,i}^{t-1}, C_g^{t-1}$ to client $\mathcal{C}_i$.

            \STATE client $\mathcal{C}_i$ computes the actual local label distribution $\left\{\&_{t'=1}^t\mathcal{Y}_{\mathcal{C}_i}^t\right\}$.

            \STATE client $\mathcal{C}_i$ computes $Y_{\mathcal{C}_i,A}^t$ through local data distribution reconstruction scheme.

            \STATE client $\mathcal{C}_i$ creates the generate replay dataset $\mathcal{D}_{A_i, C_{g,i}}^{t-1}$ with $A_i^{t-1}$ and $C_{g,i}^{t-1}$ through $Y_{\mathcal{C}_i,A}^t$. 

            \STATE client $\mathcal{C}_i$ updates $C_g^{t-1}$ on $\left\{\mathcal{D}_{A_i, C_{g,i}}^{t-1}\cup\mathcal{D}_{\mathcal{C}_i}^{t}\right\}$ by optimizing \hyperref[F_6]{$F_6$} then obtains $C_i^{t,*}$.

            \FOR{each category $c\in\left\{\&_{t'=1}^t\mathcal{Y}_{\mathcal{C}_i}^t\right\}$}
               \IF{$c\in\mathcal{Y}_{\mathcal{C}_i}^t$ \textbf{and} client $\mathcal{C}_i$ judges that $A_{i,c}^{t-1}$ need to be updated}
                  \STATE client $\mathcal{C}_i$ updates $A_{i,c}^{t-1}$ on $\mathcal{D}_{\mathcal{C}_i,y=c}^{t}$ by optimizing \hyperref[F_7]{$F_7$} then obtains $A_{i,c}^{t,*}$.
                  \STATE client $\mathcal{C}_i$ regards $A_{i,c}^{t,*}$ as $A_{i,c}^{t}$.
               \ELSE   
                  \STATE client $\mathcal{C}_i$ regards $A_{i,c}^{t-1}$ as $A_{i,c}^{t}$ without updating model.
               \ENDIF   
            \ENDFOR

            client $\mathcal{C}_i$ sends $C_i^{t,*}$, $\left\{A_{i,c}^{t,*}\right\}$ and $\left\{\&_{t'=1}^t\mathcal{Y}_{\mathcal{C}_i}^t\right\}$ to the server.
         \ENDFOR

         \STATE // Server aggregating
         \FOR{each cilent $\mathcal{C}_i\in\bm{\mathcal{C}}$}

            \STATE server updates the auxiliary model cache $A_i^{t-1}$ with $\left\{A_{i,c}^{t,*}\right\}$ then synchronizes $A_i^{t}$.

            \STATE server creates generate replay dataset $\mathcal{D}_{A_i, C_i^*}^{t}$ with $A_i^t$ and $C_i^{t,*}$ through $\left\{\&_{t'=1}^t\mathcal{Y}_{\mathcal{C}_i}^t\right\}$.

            \STATE server optimizes \hyperref[F_8]{$F_8$} on $\mathcal{D}_{A_i, C_i^*}^{t}$ then obtains personalized aggregated weights ${\bm{W}}_{i}^{t,*}=\left\{{w}_{i,j}^{t,*}\right\}_{j=1}^n$.

            \STATE server aggregates personalized global task model $C_{g,i}^t\leftarrow\sum_{j=1}^{n}({w}_{i,j}^{t,*}C_j^{t,*})$
         \ENDFOR

         server aggregates global task model $C_{g}^t\leftarrow\frac{1}{n}\sum_{i=1}^{n} C_i^{t,*}$
      \ENDFOR

   \end{algorithmic}
\end{algorithm}

\clearpage

\subsection{Evaluation Metrics}
\label{Appendix C.2}

We evaluate the performance of each method based on Instant Average Accuracy (IAA), Average Accuracy (AA) and Average Forgetting Measure (AFM). Assuming the client set is $\bm{\mathcal{C}}$ and the total number of FL rounds is $T$, the definitions of the above metrics are as follows:

\textbf{Instant Average Accuracy}. After global aggregation in each FL round $t$, we evaluate the performance of the task models on all test data corresponding to previous $t$ tasks on each client $\mathcal{C}_i\in\bm{\mathcal{C}}$ (i.e. accuracy, denoted as $a_i^t$), then calculate the IAA value of the $t$-th FL round based on the weighted average of the total number of training data encountered by each client $\mathcal{C}_i$ (denoted as $n_i^t$):
\begin{equation}
  {IAA}^t = \frac{1}{\sum_{\mathcal{C}_i \in \bm{\mathcal{C}}} n_i^t} 
  \sum_{\mathcal{C}_i \in \bm{\mathcal{C}}} n_i^t \cdot a_i^t
\end{equation}
IAA can indicate the comprehensive performance of the task model obtained in a certain FL round $t$ on all previous tasks.

\textbf{Average Accuracy}. This metric indicates the average performance of each method over the entire FL process based on the mean of the IAA values of all $T$ FL rounds: 
\begin{equation}
  AA = \frac{1}{T} \sum_{t=1}^T {IAA}^t
\end{equation}
AA can reduce the evaluation error caused by the changes of the task difficulty, and better evaluate the performance stability of different FL methods throughout the entire FL process.

\textbf{Average Forgetting Measure}. We define the forgetting measure as the difference in the performances of the client when it can access real data of previous tasks and when it cannot access real data of previous tasks. Defining the IAA value of the Centralized method in the $t$-th FL round as ${IAA}^{t}_{Centralized}$, the average forgetting measure (AFM) of each method is the average of the forgetting measure of the entire FL process:
\begin{equation}
  AFM = \frac{1}{T} \sum_{t=1}^T ({IAA}^{t}-{IAA}^{t}_{Centralized})
\end{equation}
AFM can evaluate the degree of knowledge backward transfer, and the smaller the value, the better the memory stability of the FL method.

\subsection{Detailed of Experimental Setup}
\label{Appendix C.3}

For the task model, we choose ResNet20 \cite{46} as the task model for all FL methods except FedCIL. The local training epochs are uniformly set to 20, the optimizers are SGD, the learning rates are set to 0.01, the momentums are set to 0.9, and the weight decays are set to 0.01. For FedCIL, the task model is ACGAN model, and it use its default settings corresponding to each dataset, the local training epochs are 400.

For the auxiliary model, when the pFedGRP client judges that the auxiliary sub model needs to be trained, it performs 200 epochs of training for each category's WGAN-GP\_16 model on the MNIST series dataset, 1000 epochs of training for each category's WGAN-GP\_64 model on the Cifar series dataset, and 6000 epochs of training for each category's DDPM model on the Cifar series dataset. For the local flow model of AF-FCL, each client performs 100 epochs of local training. For TARGET and MFCL, the server performs 100 epochs of training on the auxiliary model after aggregating the global task model.

For the fine-tuning epochs of global aggregation on server, pFedGRP performs 20 epochs of personalized aggregation weight optimization for each client, FedCIL performs 100 epochs of model distillation on the global ACGAN model, other FL methods do not have a fine-tuning stage for global aggregation.

\subsection{Detailed of Calculation and Communication Cost}
\label{Appendix C.4}

\begin{figure}[ht]
   \vskip -0.1in
   \begin{center}
   \caption{\text{\small\sl Table 4.} FLOPs and Parameter of the models.}
   \label{Table 4}
   \vskip 0.15in
   \centerline{\includegraphics[width=0.65\columnwidth]{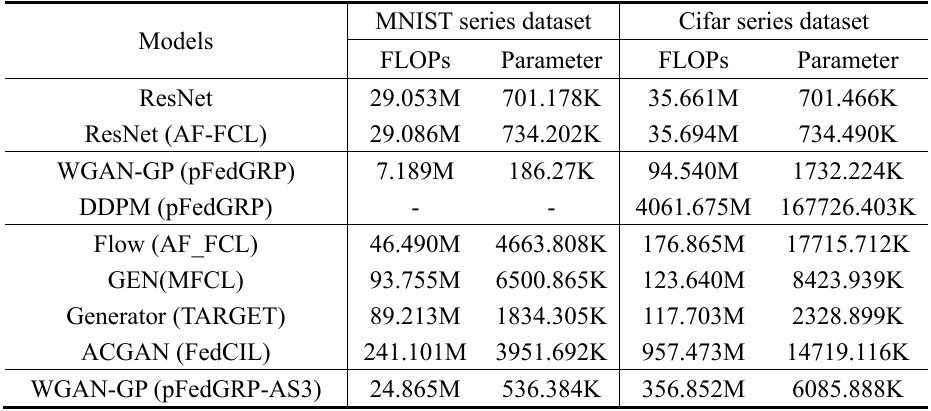}}
   \end{center}
   \vskip -0.2in
\end{figure}

Tables \hyperref[Table 4]{4} shows the FLOPs and the Parameter of all models used in our experiment. 

\begin{figure}[ht]
   \vskip -0.1in
   \begin{center}
   \caption{\text{\small\sl Table 5.} The average local additional cost on MNIST and FashionMNIST datasets}
   \label{Table 5}
   \vskip 0.15in
   \centerline{\includegraphics[width=0.75\columnwidth]{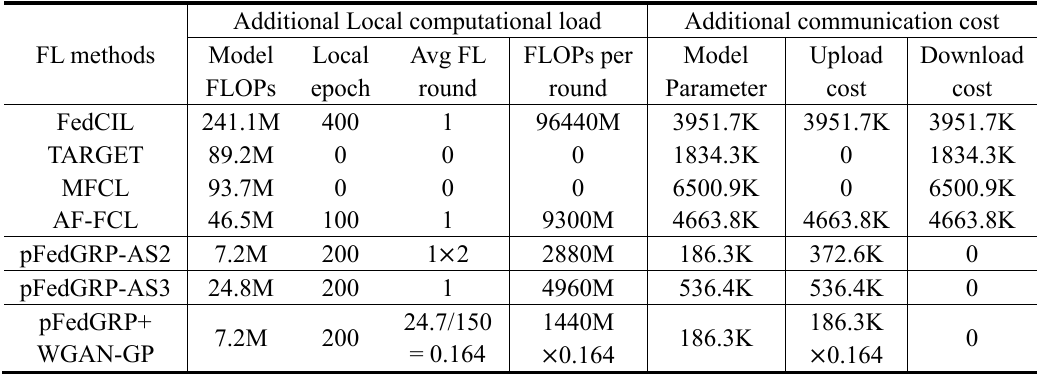}}
   \end{center}
   \vskip -0.2in
\end{figure}

\begin{figure}[ht]
   \vskip -0.1in
   \begin{center}
   \caption{\text{\small\sl Table 6.} The average local additional cost on Cifar10 dataset}
   \label{Table 6}
   \vskip 0.15in
   \centerline{\includegraphics[width=0.75\columnwidth]{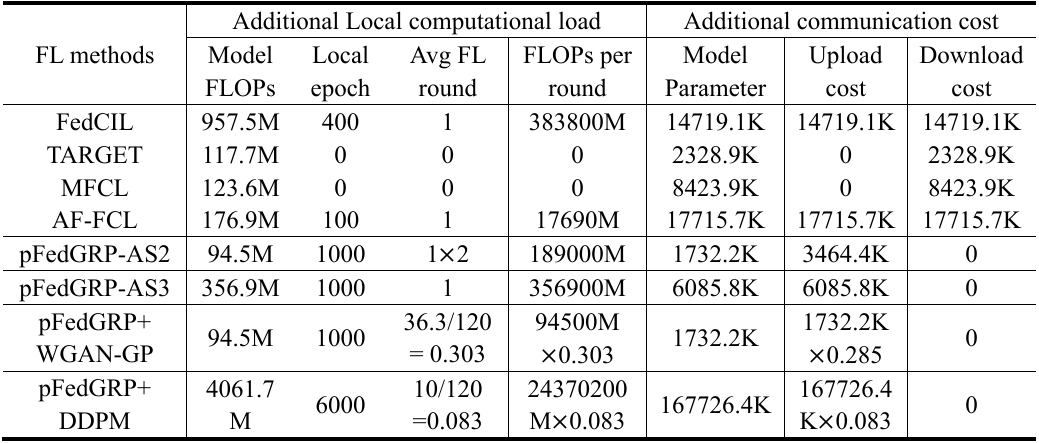}}
   \end{center}
   \vskip -0.6in
\end{figure}

In our experiment, clients of pFedGRP train the WGAN-GP model in an average of 24.7 times in the total 150 FL rounds of the MNIST and FashionMNIST datasets, train the WGAN-GP model an average of 36.3 times and train the DDPM model an average of 10 times in the total 120 FL rounds of the Cifar10 dataset. Tables \hyperref[Table 5]{5} and Table \hyperref[Table 6]{6} show the average local additional computational load and additional communication cost of each FL round which is bring by training the auxiliary model.

\clearpage

\section{Additional Experimental Results}
\label{Appendix D}
\subsection{Ablation Experiments}
\label{Appendix D.1}

pFedGRP framework mainly consists of generation replay portion and federation portion. In two scenarios of baseline experiments constructed on the MNIST, FMNIST, and Cifar10 datasets, we conducted ablation studies on each point of the two portions. The auxiliary sub models used in ablation experiments are all WGAN-GP.

For the generated replay portion, we will conduct ablation study from the following points: 

1. pFedGRP no longer uses task models to select the generating data of the auxiliary sub models, which is denoted as pFedGRP-AS1, and the quality of generated replay may decrease to a certain extent. 

2. pFedGRP no longer uses local task model to determine whether auxiliary sub models need to be updated, but updates auxiliary sub models in each FL round, which is denoted as pFedGRP-AS2, then the computational and communication costs of updating the auxiliary model on the client side will significantly increase. 

3. pFedGRP combines with the generating replay scheme of other FCL methods, each client uses a single WGAN-GP model with double channels (32 channels for the MNIST series dataset and 128 channels for the Cifar10 dataset) as auxiliary model, which is denoted as pFedGRP-AS3. At each epoch of local training and global aggregation, this method uses auxiliary model to generate data of random categories whose soft labels are determined by the task model obtained previously, and the local auxiliary model will be updated on real data and its own generated data in each FL round.

For the federation portion, we will conduct ablation study from the following points: 

1. pFedGRP only uses the global task model to initialize the local task model, but no longer aligns the output of the local task model and the personalized global task model on the generated data, which is denoted as pFedGRP-ASG, where the local task model is only trained with hard labels. 

2. Furthermore, pFedGRP only uses the personalized global task model to initialize the local task model, which is denoted as pFedGRP-ASP, then the local task model will contain less global information. 

3. Combining the classic FL method FedAVG and personalized FL method pFedGraph with the generated replay portion of pFedGRP, which are denoted as FedAVG-replay and pFedGraph-replay, thus verifying the performance of the federation portion of pFedGRP.

\begin{figure}[ht]
   \vskip -0.1in
   \begin{center}
   \caption{\text{\small\sl Table 7.} Ablation Study Results on FL with Tasks Gradually Changing}
   \label{Table 7}
   \vskip 0.15in
   \centerline{\includegraphics[width=0.65\columnwidth]{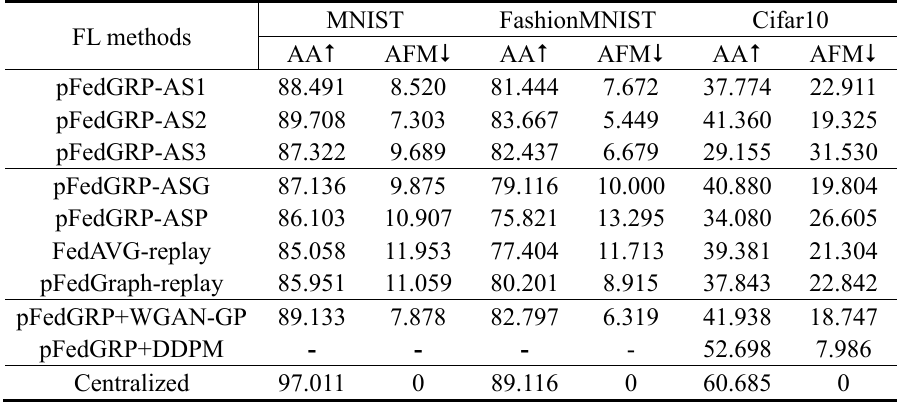}}
   \end{center}
   \vskip -0.2in
\end{figure}

The experimental results of the seven ablation methods mentioned above and the pFedGRP method are shown in Table \hyperref[Table 7]{7} and Table \hyperref[Table 8]{8}. The IAA variation of all methods above and corresponding analysis are shown in Appendix \hyperref[Appendix E.4]{E.4}, and the calculation and communication consumption of all FL methods above are shown in Appendix \hyperref[Appendix C.4]{C.4}.

Furthermore, we calculated the FID values \cite{47} of the generated replay schemes used by various methods in the ablation study. The lower the value, the better the performance of the generated replay. The final results are shown in Table \hyperref[Table 9]{9} below. It can be seen that as the complexity of data increases, the generated replay effect of the auxiliary model with category decoupling gradually becomes much better than that of a single larger auxiliary model. On this basis, using the information contained in the task model can further enhance the generated replay performance of the auxiliary model.
\begin{figure}[ht]
   \vskip -0.1in
   \begin{center}
   \caption{\text{\small\sl Table 8.} Ablation Study Results on FL with Tasks Circulating}
   \label{Table 8}
   \vskip 0.15in
   \centerline{\includegraphics[width=0.65\columnwidth]{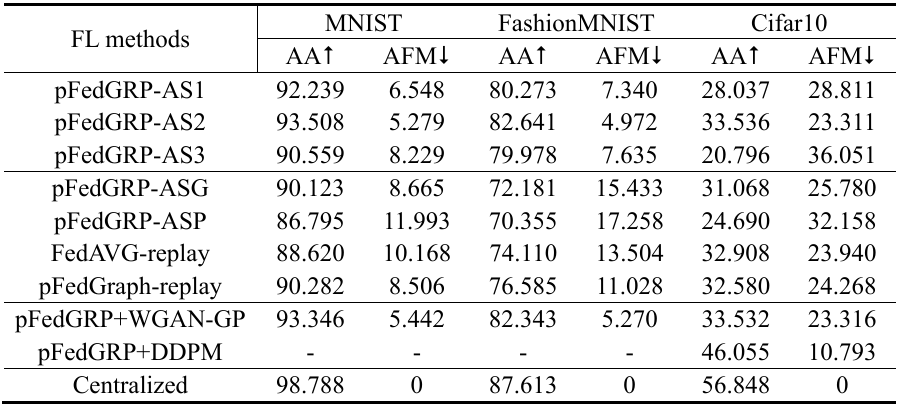}}
   \end{center}
   \vskip -0.2in
\end{figure}

\begin{figure}[ht]
   \vskip -0.1in
   \begin{center}
   \caption{\text{\small\sl Table 9.} FID values for various Generated Replay Schemes}
   \label{Table 9}
   \vskip 0.15in
   \centerline{\includegraphics[width=0.65\columnwidth]{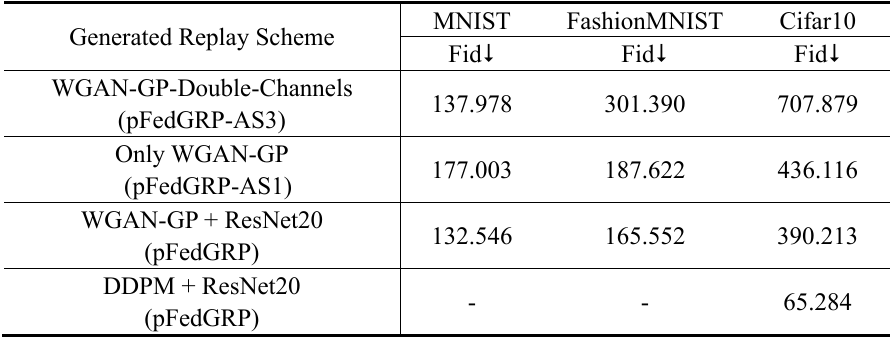}}
   \end{center}
   \vskip -0.2in
\end{figure}

\subsection{Baseline Experiments on FL with Different Correlations Between Tasks}
\label{Appendix D.2}

\begin{figure}[ht]
  \vskip -0.1in
  \begin{center}
  \caption{\text{\small\sl Table 10.} Baseline Experiment Results on MNIST and FL with Tasks Gradually Changing}
  \label{Table 10}
  \vskip 0.15in
  \centerline{\includegraphics[width=0.75\columnwidth]{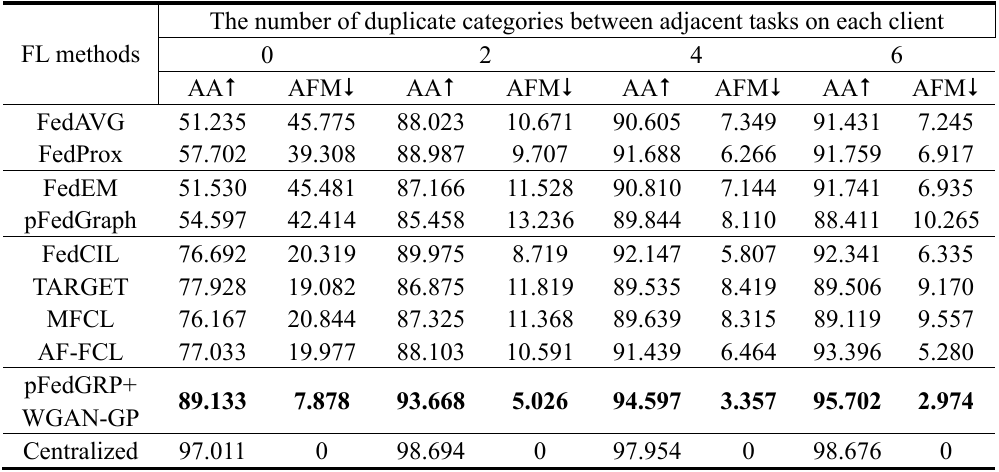}}
  \end{center}
  \vskip -0.2in
\end{figure}

\begin{figure}[ht]
  \vskip -0.1in
  \begin{center}
  \caption{\text{\small\sl Table 11.} Baseline Experiment Results on FashionMNIST and FL with Tasks Gradually Changing}
  \label{Table 11}
  \vskip 0.15in
  \centerline{\includegraphics[width=0.75\columnwidth]{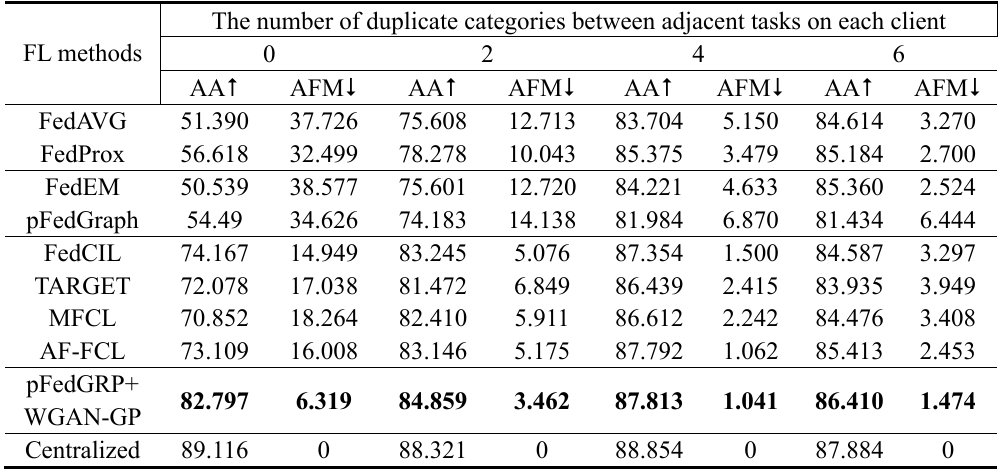}}
  \end{center}
  \vskip -0.2in
\end{figure}

\begin{figure}[ht]
  \vskip -0.1in
  \begin{center}
  \caption{\text{\small\sl Table 12.} Baseline Experiment Results on Cifar12 and FL with Tasks Gradually Changing}
  \label{Table 12}
  \vskip 0.15in
  \centerline{\includegraphics[width=0.75\columnwidth]{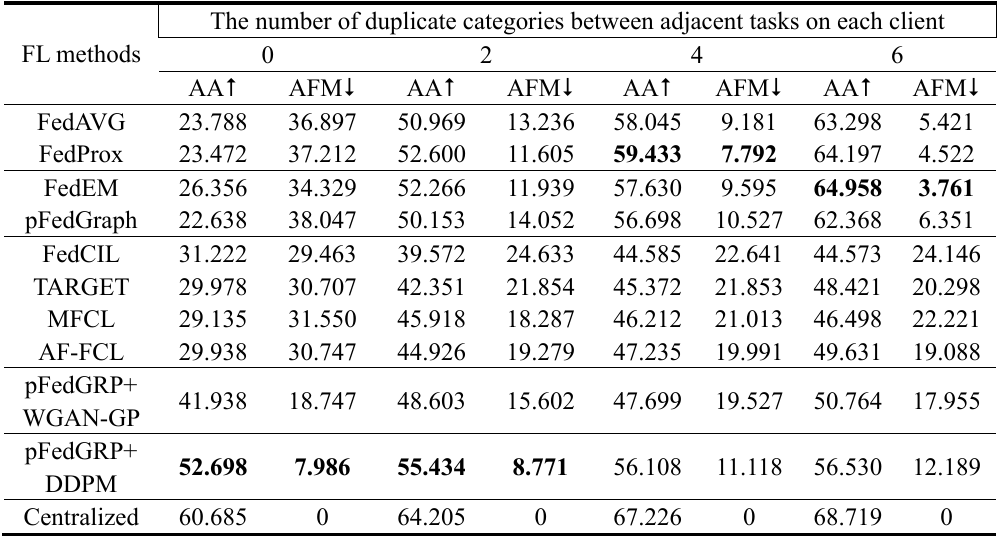}}
  \end{center}
  \vskip -0.2in
\end{figure}

On the setting of the first baseline experiments (i.e. FL with Tasks Gradually Changing), We further investigated the performance changes of pFedGRP and various FL baseline methods when the correlation between tasks is gradually increasing. 
Since the number of duplicate categories between adjacent tasks of each client in the baseline setting is 0, we increased this number to 2, 4 and 6 (i.e. each task has 4, 6 and 8 categories respectively), and the number of real data for each category remains at 200. Due to the limited amount of real data in the dataset, as the heterogeneity of data between and within clients decreases, the total number of rounds in FL and the total number of tasks for each client decreases to 70, 50 and 30, respectively (for Cifar10 is 60, 40 and 30). The results of pFedGRP and other baseline methods in the various experimental settings mentioned above are presented in Table \hyperref[Table 10]{10}, Table \hyperref[Table 11]{11} and Table \hyperref[Table 12]{12}. 


It can be seen from the tables above that the performance improvement of all FL methods are significant with the decrease of data heterogeneity. However, on Cifar10 dataset with complex data distribution, the data distribution replayed by the auxiliary model often deviates significantly from the real data distribution, making the performance of the four FCL methods and the pFedGRP method inferior to the FL methods and the pFL methods on lower data heterogeneity. Due to the adoption of many strategies to reduce the generated replay errors, the performance of pFedGRP leads all FCL methods in all experimental settings.

\clearpage

\twocolumn

\section{IAA Variation Charts for Experiments}
\label{Appendix E}

\subsection{IAA Variation Charts for Tasks Gradually Changing}
\label{Appendix E.1}


\begin{figure}[ht]
   \vskip -0.05in
   \begin{center}
   \centerline{\includegraphics[width=0.8\columnwidth]{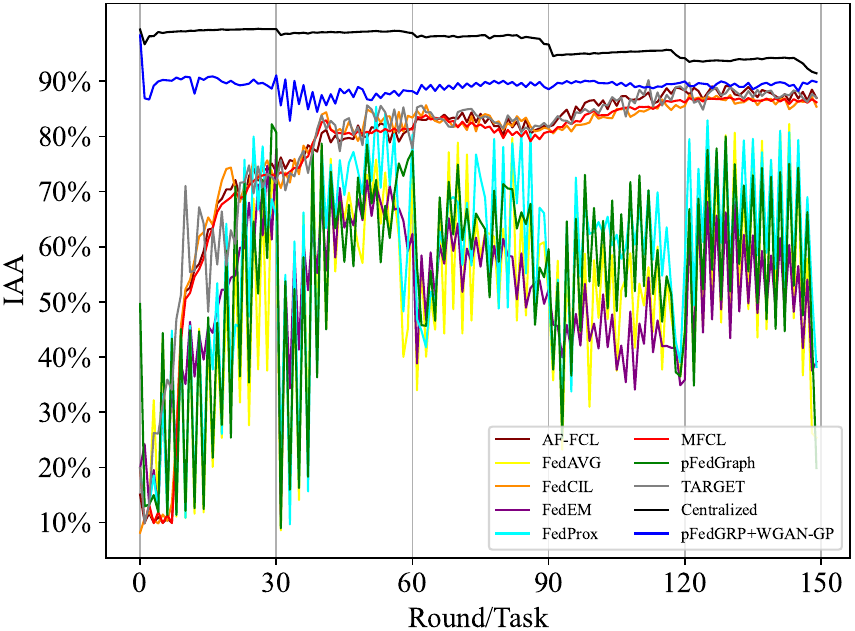}}
   \caption{\text{\small\sl Figure 6.} IAA Variation Chart of baseline experiment for Tasks Gradually Changing in MNIST dataset.}
   \end{center}
   \vskip -0.2in
\end{figure}
\begin{figure}[ht]
   \vskip -0.1in
   \begin{center}
   \centerline{\includegraphics[width=0.8\columnwidth]{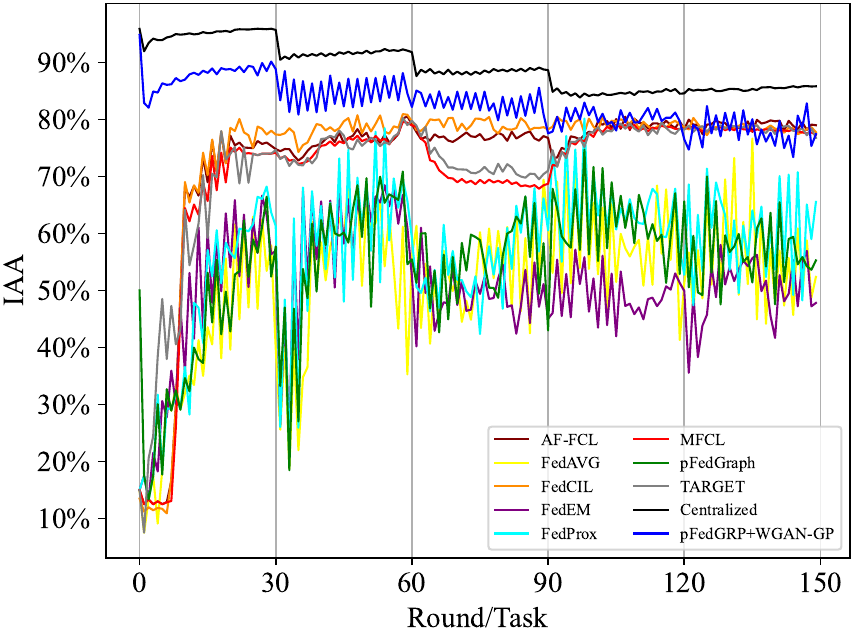}}
   \caption{\text{\small\sl Figure 7.} IAA Variation Chart of baseline experiment for Tasks Gradually Changing in FashionMNIST dataset.}
   \end{center}
   \vskip -0.2in
\end{figure}
\begin{figure}[ht]
   \vskip -0.1in
   \begin{center}
   \centerline{\includegraphics[width=0.8\columnwidth]{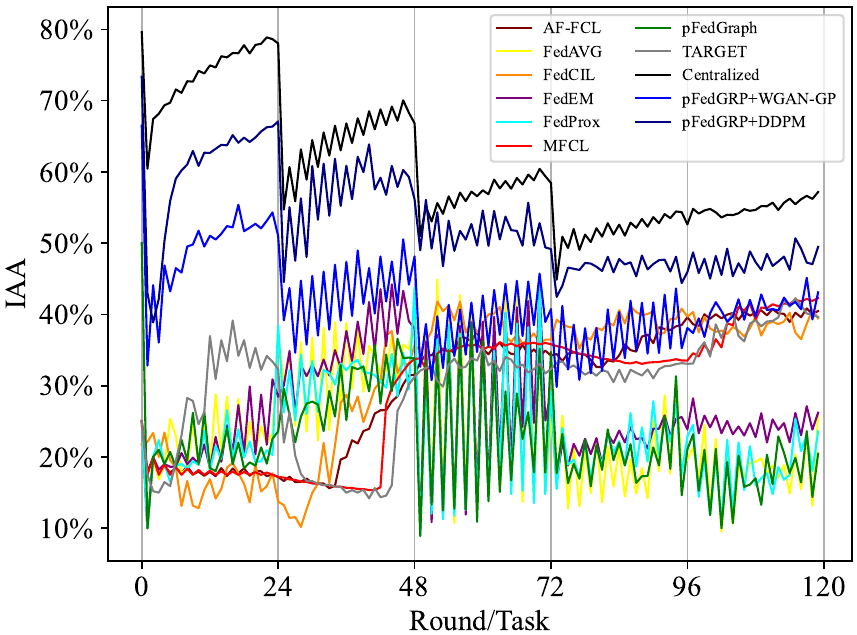}}
   \caption{\text{\small\sl Figure 8.} IAA Variation Chart of baseline experiment for Tasks Gradually Changing in Cifar10 dataset.}
   \end{center}
   \vskip -0.8in
\end{figure}

\vspace{0.6in}

Under the FL setting of Tasks Gradually Changing, the gray vertical lines in the figure correspond to the FL rounds where the types of tasks of each client's task loop changes. Overall, pFedGRP achieve good performance in the early and middle stages of FL training by effectively estimating the data distribution of each client to aggregate personalized task models for clients. However, the baseline FCL methods require to use task model to train the auxiliary model, the convergence time of FCL methods is usually proportional to the data complexity of the dataset, resulting in poor performance in the early and middle stages of training.

\subsection{IAA Variation Charts for Tasks Circulating}
\label{Appendix E.2}


\begin{figure}[ht]
   \vskip -0.05in
   \begin{center}
   \centerline{\includegraphics[width=0.8\columnwidth]{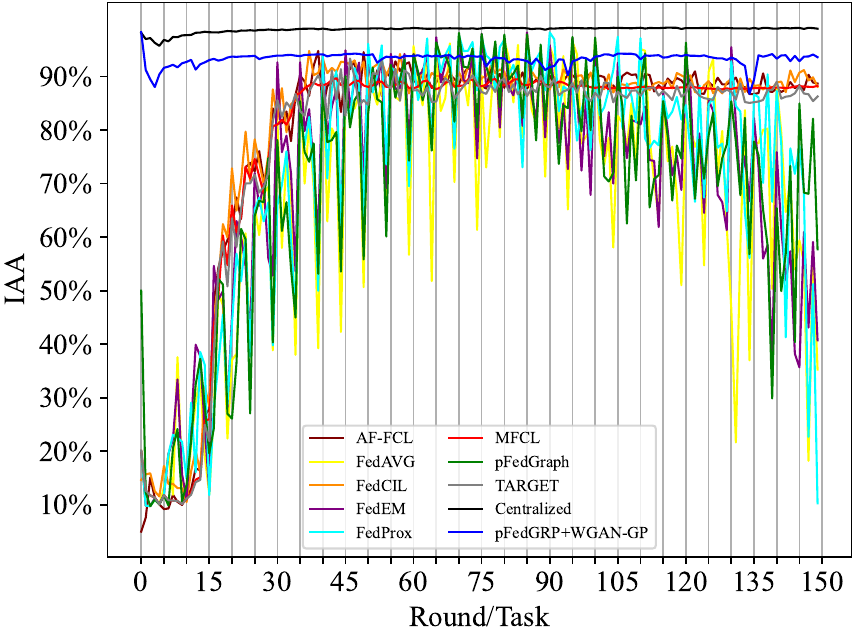}}
   \caption{\text{\small\sl Figure 9.} IAA Variation Chart of baseline experiment for Tasks Circulating in MNIST dataset.}
   \end{center}
   \vskip -0.2in
\end{figure}
\begin{figure}[ht]
   \vskip -0.1in
   \begin{center}
   \centerline{\includegraphics[width=0.8\columnwidth]{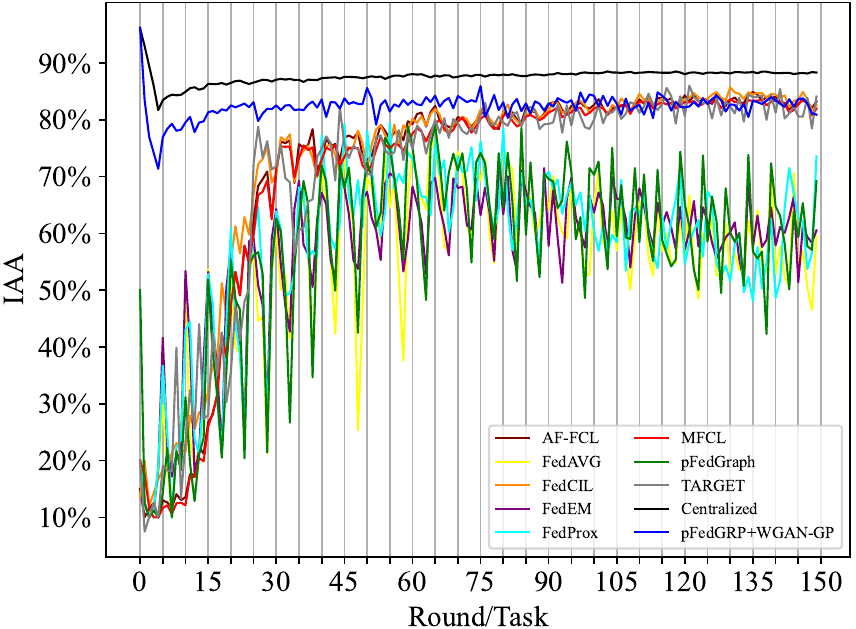}}
   \caption{\text{\small\sl Figure 10.} IAA Variation Chart of baseline experiment for Tasks Circulating in FashionMNIST dataset.}
   \end{center}
   \vskip -0.2in
\end{figure}
\begin{figure}[ht]
   \vskip -0.1in
   \begin{center}
   \centerline{\includegraphics[width=0.8\columnwidth]{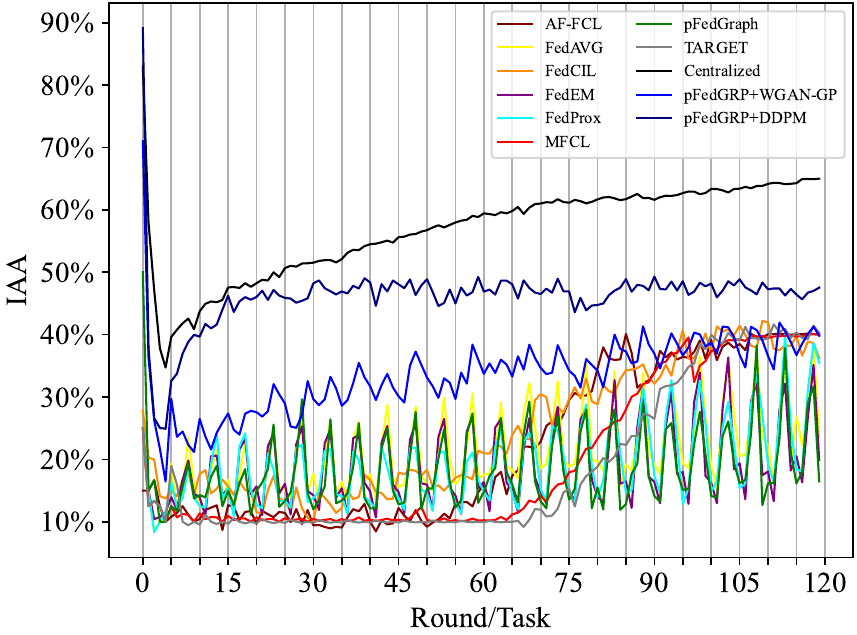}}
   \caption{\text{\small\sl Figure 11.} IAA Variation Chart of baseline experiment for Tasks Circulating in Cifar10 dataset.}
   \end{center}
   \vskip -0.8in
\end{figure}

\clearpage

Under the FL setting of Tasks Circulation, the gray vertical line in the figure corresponds to the FL round at the beginning of each task cycle on each client (i.e. five rounds), meaning that the distribution of data encountered by the client in every five rounds is similar to the data distribution of the entire FL process. The conclusion drawn from the experimental results under this setting is similar to that of the previous experiment.

\subsection{IAA Variation Charts for FL under High Data Heterogeneity}
\label{Appendix E.3}


\begin{figure}[ht]
   \vskip -0.05in
   \begin{center}
   \centerline{\includegraphics[width=0.8\columnwidth]{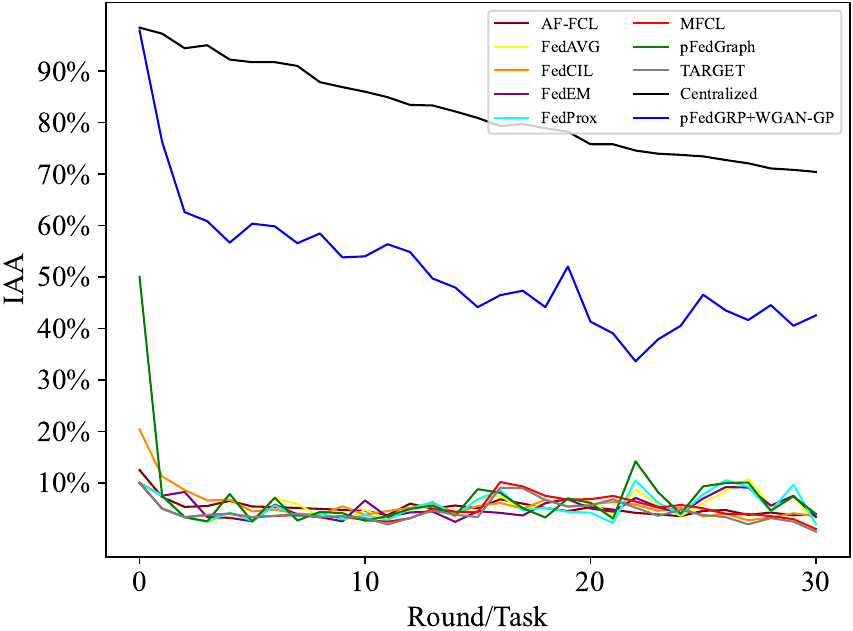}}
   \caption{\text{\small\sl Figure 12.} IAA Variation Chart of baseline experiment for High Data Heterogeneity in EMNIST-ByClass dataset.}
   \end{center}
   \vskip -0.2in
\end{figure}
\begin{figure}[ht]
   \vskip -0.1in
   \begin{center}
   \centerline{\includegraphics[width=0.8\columnwidth]{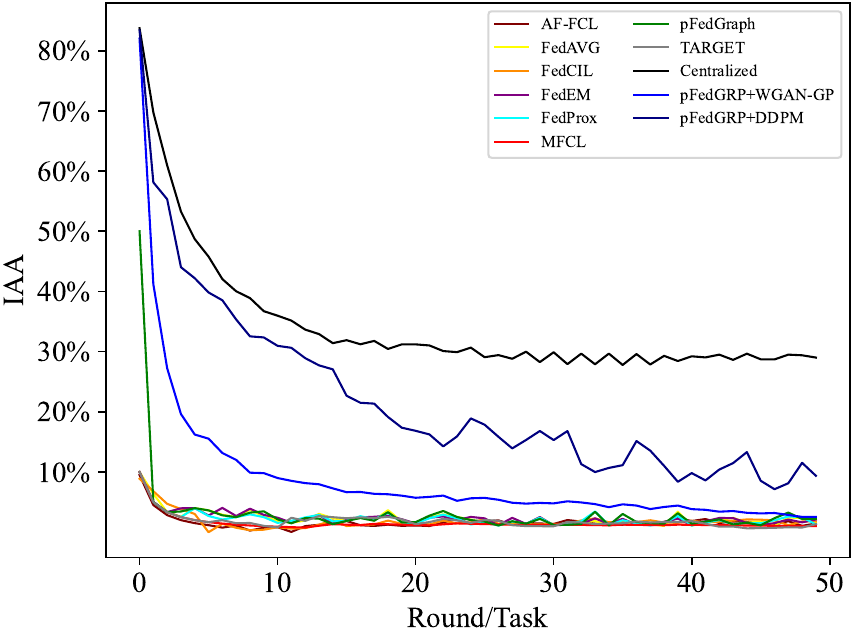}}
   \caption{\text{\small\sl Figure 13.} IAA Variation Chart of baseline experiment for High Data Heterogeneity in Cifar100 dataset.}
   \end{center}
   \vskip -0.8in
\end{figure}

\vspace{0.6in}

Under the FL setting of High Data Heterogeneity, each client will encounter two categories of data in the new FL round that they have not encountered before, until all categories in the dataset are traversed. This means that the FL setting in this experiment is similar to the one-shot FL which makes it impossible for all FL methods to converge, further testing the robustness of these FL methods. It can be seen that the pFedGRP method performs much better than other baseline methods when continuously encountering new categories.

\vspace{0.2in}

\subsection{IAA Variation Charts for Ablation Study}
\label{Appendix E.4}


\begin{figure}[ht]
   \vskip -0.05in
   \begin{center}
   \centerline{\includegraphics[width=0.8\columnwidth]{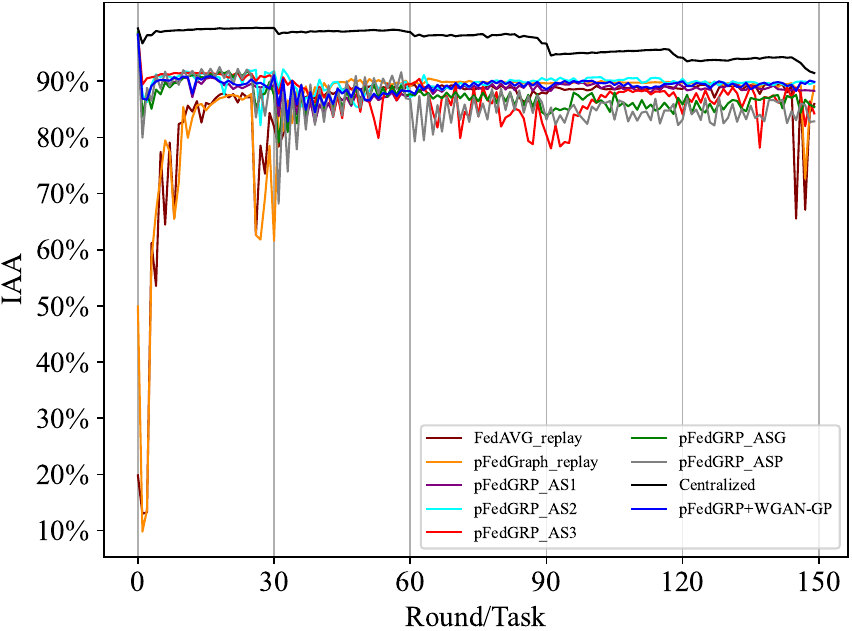}}
   \caption{\text{\small\sl Figure 14.} IAA Variation Chart of Ablation Study for Tasks Gradually Changing in MNIST dataset.}
   \end{center}
   \vskip -0.2in
\end{figure}
\begin{figure}[ht]
   \vskip -0.1in
   \begin{center}
   \centerline{\includegraphics[width=0.8\columnwidth]{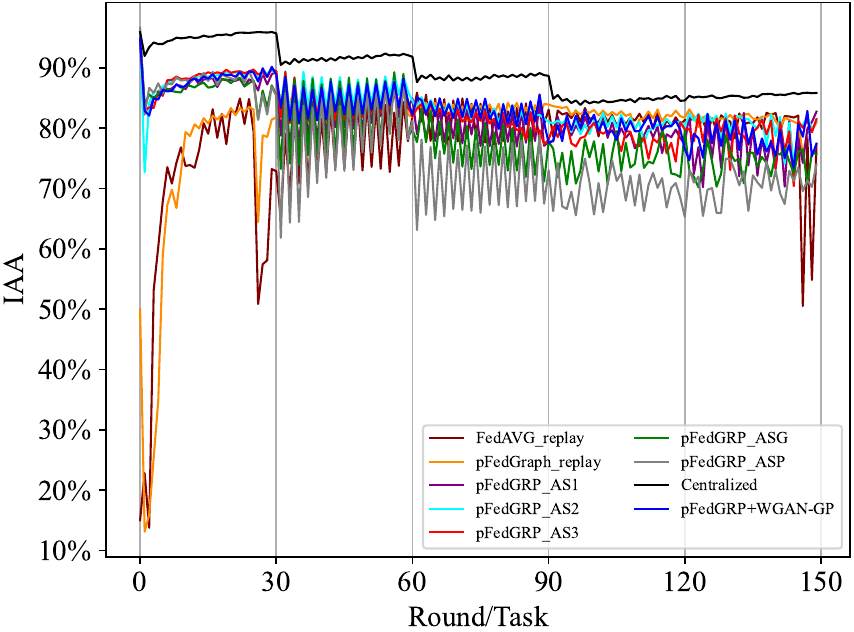}}
   \caption{\text{\small\sl Figure 15.} IAA Variation Chart of Ablation Study for Tasks Gradually Changing in FashionMNIST dataset.}
   \end{center}
   \vskip -0.2in
\end{figure}
\begin{figure}[ht]
   \vskip -0.1in
   \begin{center}
   \centerline{\includegraphics[width=0.8\columnwidth]{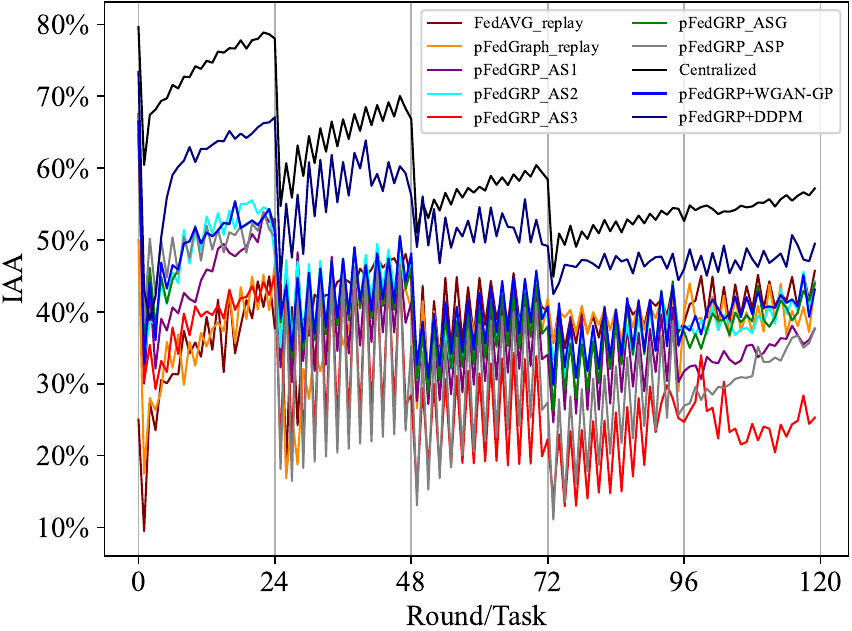}}
   \caption{\text{\small\sl Figure 16.} IAA Variation Chart of Ablation Study for Tasks Gradually Changing in Cifar10 dataset.}
   \end{center}
   \vskip -2in
\end{figure}

\clearpage

\begin{figure}[ht]
   \vskip 0.2in
   \begin{center}
   \centerline{\includegraphics[width=0.8\columnwidth]{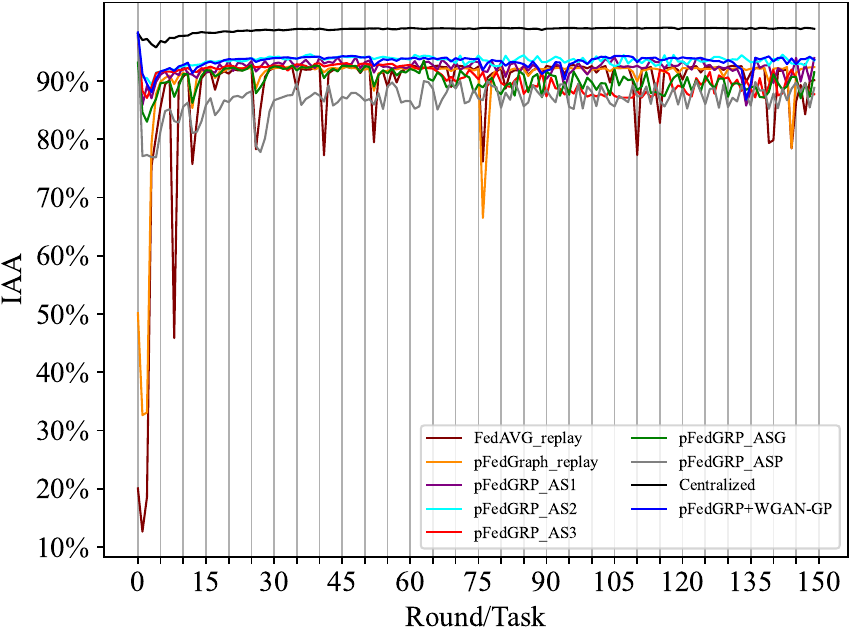}}
   \caption{\text{\small\sl Figure 17.} IAA Variation Chart of Ablation Study for Tasks Circulating in MNIST dataset.}
   \end{center}
   \vskip -0.2in
\end{figure}
\begin{figure}[ht]
   \vskip -0.1in
   \begin{center}
   \centerline{\includegraphics[width=0.8\columnwidth]{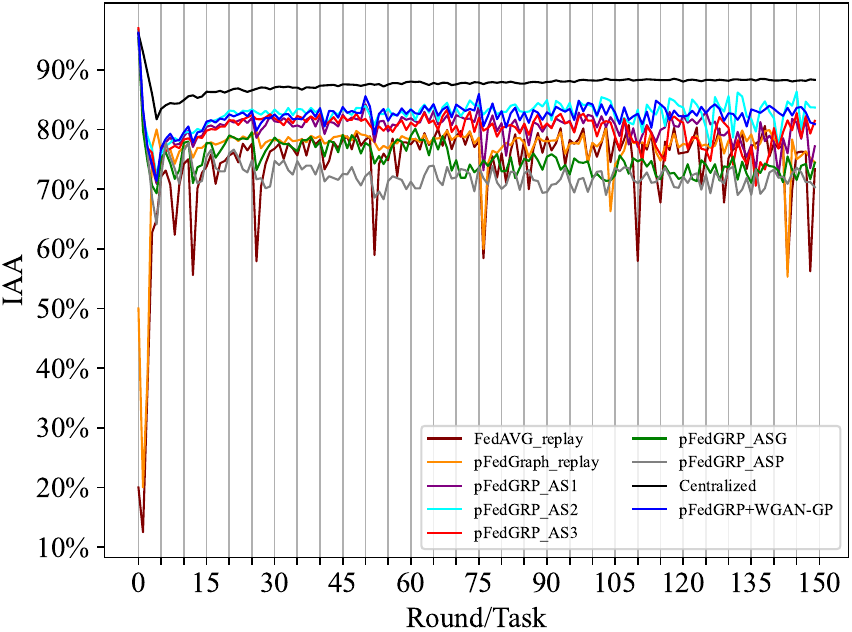}}
   \caption{\text{\small\sl Figure 18.} IAA Variation Chart of Ablation Study for Tasks Circulating in FashionMNIST dataset.}
   \end{center}
   \vskip -0.2in
\end{figure}
\begin{figure}[ht]
   \vskip -0.1in
   \begin{center}
   \centerline{\includegraphics[width=0.8\columnwidth]{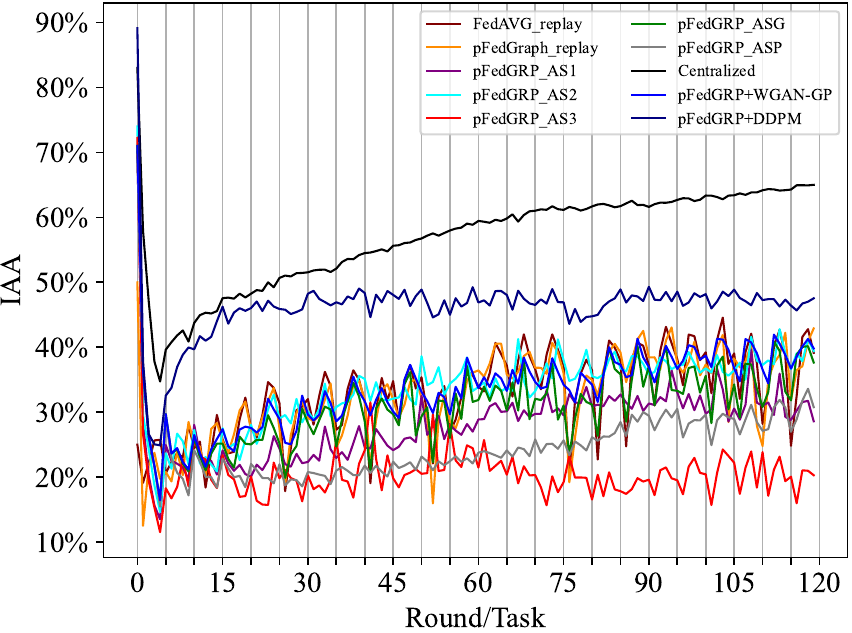}}
   \caption{\text{\small\sl Figure 19.} IAA Variation Chart of Ablation Study for Tasks  Circulating in Cifar10 dataset.}
   \end{center}
   \vskip -2in
\end{figure}

\vspace{1.8in}

The following points can be seen from the figures above: 

1. The performance of pFedGRP-AS1 is inferior to that of pFedGRP in all scenarios, indicating that using task models to select generated data can effectively reduce replay errors. 

2. pFedGRP-AS2 updates the auxiliary sub models in each FL round, but its performance is only slightly higher than that of pFedGRP, indicating that the necessity of updating the auxiliary model in each FL round is not high. 

3.With the generate replay scheme of other FCL methods, pFedGRP-AS3 achieves the worst performance with a huge amount of computation, proving the efficiency of the generated replay scheme of pFedGRP. 

4. Without using the local knowledge transfer scheme of pFedGRP, the performance of the pFedGRP-ASG, which uses the global task model to initialize the local task model, is inferior to that of pFedGRP, but this gap decreases as the complexity of the dataset increases, which means that local knowledge transfer can alleviate model forgetting to some extent. 

5. Without using the global task model to initialize the local task model, the pFedGRP-ASP method, which uses personalized global task model to initialize the local task model, performs much worse than pFedGRP and pFedGRP-ASG in the later stages of FL training, meaning that using a global task model to initialize a local task model can improve the generalization ability of task model. 

6. Without using the personalized aggregation scheme of pFedGRP, FedAVG-replay and pFedGraph-replay performs worse than pFedGRP in the later stages of FL training, but their performance are similar to that of pFedGRP in the middle and later stages of FL training, meaning that pFedGRP can more effectively address the complex data heterogeneity between clients.

\end{document}